\documentclass[trackchanges,twocolumn,twocolappendix,times]{aastex701}

\newcommand{\muhz}{\,$\mu$Hz} 
\newcommand{\err}[2]{\ensuremath{^{+#2}_{-#1}}}
\newcommand{\rej}{RE J1034+396}

\usepackage{graphicx}	
\usepackage{amsmath}	
\usepackage{upgreek}

 \shorttitle{Periodic Behaviour in NGC~6814}
 \shortauthors{Hodd et al.}

\begin{document}

\title{Investigating the Periodic X-ray Behaviour in the Eclipsing AGN NGC~6814}

\author[orcid=0009-0000-5162-3185,gname=Thomas, sname=Hodd]{T. O. Hodd}
\affiliation{Department of Astronomy and Physics, Saint Mary’s University, 923 Robie Street, Halifax, NS B3H 3C3, Canada}
\email[show]{thomas.hodd@smu.ca}

\author[orcid=0009-0006-4968-7108,gname=Luigi, sname=Gallo]{L. C. Gallo} 
\affiliation{Department of Astronomy and Physics, Saint Mary’s University, 923 Robie Street, Halifax, NS B3H 3C3, Canada}
\email{luigi.gallo@smu.ca}

\author[orcid=0000-0003-3678-5033,gname=Adam,sname=Gonzalez]{A. G. Gonzalez}
\affiliation{Department of Astronomy and Physics, Saint Mary’s University, 923 Robie Street, Halifax, NS B3H 3C3, Canada}
\email{adam.gonzalez@smu.ca}

\author[orcid=0000-0003-2869-7682,gname=Jon,sname=Miller]{J. M. Miller}
\affiliation{Department of Astronomy, University of Michigan, 1085 South University Avenue, Ann Arbor, MI 48109-1107, USA}
\email{jonmm@umich.edu}

\author[orcid=0000-0002-5359-9497,gname=Daniele,sname=Rogantini]{D. Rogantini}
\affiliation{Department of Astronomy and Astrophysics, University of Chicago, 5640 S Ellis Avenue, Chicago, IL 60637, USA}
\email{danieler@uchicago.edu}

\begin{abstract}
A 2016 \textit{XMM-Newton} X-ray light curve of the Seyfert 1.5 galaxy NGC~6814 exhibited clear eclipsing behaviour, with distinct ingress and egress, during half of the observation. Here, we report on the periodic behaviour in the light curve prior to the eclipse. We use timing and spectral analysis techniques to quantify the behaviour and examine the characteristics of the periodic signal. A superlet transform of the X-ray light curve reveals a period of $\sim45-50$\muhz\ in the initial $60$\,ks of the observation with a detection significance at the \textgreater90\% level in both the broad ($0.3-10$\,keV) and soft ($0.3-1.0$\,keV) bands. The period is confirmed by fitting a sinusoid, and is also evident in the highest energy bands with diminished significance because of reduced signal-to-noise. There appear to be distinct changes in the variability behaviour during the eclipse as the measured period is modulated (stretched) at all energies. From phase-resolved spectra, we investigate possible physical causes of this periodic behaviour and find that it can be interpreted as changes in the covering fraction or a non-standard inner accretion flow (e.g. truncated disc and misaligned flow). The non-standard inner flow appears consistent with previous reports of a truncated inner disc and compact corona in NGC~6814.

\end{abstract}



\section{Introduction}
Active galactic nuclei (AGN) are amongst the most luminous X-ray objects in the Universe. The central region around the supermassive black hole (SMBH) is so compact it is unresolved by current telescopes apart from two isolated cases observed by the \textit{Event Horizon Telescope} (EHT): M87 \citep{eht2019}, and Sgr A* \citep{eht2022}. Instead, we can utilise timing or spectral analysis techniques to investigate the X-ray emitting region. Close to the SMBH, the hot corona illuminates the inner part of the accretion disc \citep[e.g.][]{fabian2009}. This hot corona up-scatters the UV photons emitted by the disc through Comptonisation to produce the X-ray power law we observe \citep[e.g.][]{laha2025}. An excess in the soft X-rays can be explained as either blurred reflection of the primary emission off the inner disc \citep[e.g.][]{fabian2009, fabian2010}, emission from a warm corona above the disc \citep[e.g.][]{middleton2011, done2012, petrucci2018}, or some combination of both \citep[e.g.][]{ballantyne2020}. The continuum emission can also be subject to absorption from winds and material in the line-of-sight and emission from the broad-line region (BLR) or distant torus \citep[e.g.][]{gallo2023}.

The compact nature of the inner region means that X-ray variability can be observed on timescales from hours \citep{vaughan2003} to years \citep{markowitz2001}. Studies of this variability can help explain the physical process that drive AGN, and reveal the structure surrounding the SMBH \citep[e.g.][]{vaughan2003, uttley2014}. Fourier analysis techniques such as the power spectrum (PSD), and lag-frequency/lag-energy spectra are common applications for probing the accretion disc, corona, and the black hole itself \citep[e.g.][]{fabian2009, zoghbi2010, gonzalez2025, paolillo2025}.

Periodic behaviour in the X-ray has been observed in several AGN, most significantly in \rej\ where the quasi-periodic oscillation (QPO) is strong, persistent and  well-studied \citep[e.g.][]{gierlinski2008, middleton2009}. In \rej, the QPO dominates the X-ray variability but there is debate about its origin and it has been proposed that it arises from periodic obscuration by a warm absorber \citep{maitra2010} or from within the hot corona \citep{taylor2025}. Other hypotheses for the origin of the QPO are shock oscillations in the accretion flow \citep{czerny2012}, or from super-Eddington changes in disc structure \citep{middleton2009}. Significant detections of X-ray QPOs in other AGN are rare, but QPOs are seen in other objects such as X-ray binaries \citep[e.g.][]{remillard2006}. General relativistic magnetohydrodynamic (GRMHD) simulations of tilted and truncated discs have shown that these disc geometries can induce QPO-like variability \citep{bollimpalli2024}.

NGC~6814 is a type 1.5 Seyfert galaxy at $z=0.00521$. The supermassive black hole (SMBH) mass is estimated at $\sim1.4\,\times\,10^{7}\,\mathrm{M}_{\odot}$ \citep{bentz2015}. The coronal temperature has been measured to be $45\err{17}{100}$\,keV \citep{tortosa2018}. It has been observed to have moderate absorption \citep[e.g.][]{gallo2021, walton2013, waddell2020} and variability on timescales of hours \citep[e.g.][]{walton2013} to years \citep[e.g.][]{mukai2003}. \cite{gallo2021} found an X-ray eclipse lasting 42\,ks in the 2016 \textit{XMM-Newton} observation, and determined that the obscuring material was located $\sim2700\,r_g$, consistent with broad-line region (BLR) clouds. Additionally, the X-ray corona was estimated to be $\sim25\,r_g$ in diameter, which is consistent with coronal measurements made through reverberation \citep[e.g.][]{alston2020}. The obscuring material in the eclipse is likely clumpy and inhomogeneous \citep{pottie2023, kang2023}, with such broad line clouds covering approximately $2-4\%$ of the orbital path-length for a Keplerian orbit.

In this work we report on the periodic behaviour in the light curve, and use spectral analysis techniques to determine possible physical explanations for this behaviour. In Section \ref{sec:obs} we discuss how we obtain light curves and spectra from the observation. In Section \ref{sec:timing} we explain the wavelet and superlet transforms and apply them to our data in order to search for periodic behaviour. In Section \ref{sec:mod} we explore if and how changes in density can modulate oscillation periods in AGN light curves. Then in Section \ref{sec:spectral} we use our findings from the superlet transform to inform a phase-resolved spectral analysis to explain the possible physical origins of the oscillations. In Section \ref{sec:discussion} we discuss these findings in the broader context of periodic behaviour and quasi-periodic oscillations in AGN and compare our findings with known QPOs. We give our conclusions to this work in Section \ref{sec:conclusions}.

\section{Observations and Data Reduction}
\label{sec:obs}
Our data comes from the 2016 \textit{XMM-Newton} \citep{xmm2001} observation of NGC~6814, in which an eclipsing event was seen \citep{gallo2021}. The observation lasted approximately 131\,ks starting on 6\textsuperscript{th} April 2016. In this observation, the EPIC pn camera \citep{xmmpn2001} was operated in large window mode with the medium filter. We then processed the \textit{XMM-Newton} Observation Data Files (ODFs) using the \textit{XMM-Newton} Science Analysis System (\textsc{sas v21.0.0}) to produce our calibrated event lists.

The event list was extracted from a 35'' radius circular region centred on the source, with background from a circular region (50'') away from the source but on the same CCD. We checked that pileup was negligible and extracted light curves using \textsc{evselect} in five energy bands referred to as the soft ($0.3-1.0$\,keV), mid ($1.0-4.0$\,keV), hard ($4.0-12.0$\,keV), very hard ($6.0-12.0$\,keV), and broad ($0.3-12.0$\,keV) bands. These light curves span the entire observation, excluding the last $\sim10$\,ks due to high background flaring, and were corrected for exposure with \textsc{epiclccor}. Additionally, we extracted sixteen narrow band light curves over the $0.3-12.0$\,keV range for use in our lag-energy spectra. All light curves are binned into evenly spaced 100\,s bins. Our phase-resolved spectra were extracted using good time intervals (GTIs) identified from a sinusoidal fit to the broad-band light curve (see Section \ref{sec:timing}). We generated EPIC response matrices using the \textsc{rmfgen} and \textsc{arfgen} tools.

We adopt a Galactic column density value of $1.48\times10^{21}\,\mathrm{cm}^{-2}$ \citep{willingale2013}, and use abundances based on \cite{wilms2000}. Spectra were optimally binned \citep{kaastra2016} with \textsc{ftgrouppha} and fit using \textsc{xspec} v12.14.1 \citep{arnaud1996}. The fit statistic used for all our spectral modelling was the C-Statistic \citep{cash1979}. To compare the quality of fit of different models on the same data we used the Akaike Information Criterion \cite[AIC, ][]{akaike1974}. The AIC estimates model quality based on the number of free parameters and the value of the log-likelihood function, improving with goodness of fit but penalising increases in the number of parameters.

\section{Timing Analysis}
\label{sec:timing}
\subsection{The Superlet Transform of NGC~6814}
One disadvantage of using Fourier analysis methods is the implicit assumption of a stationary process in the time series, which may not be the case for the X-ray light curves of AGN \citep{alston2019}. Although modifications can be made to achieve some time resolution in the Fourier transform (e.g. the short-time Fourier transform), the wavelet transform naturally preserves time domain information by using wave packets (wavelets) localised in time instead of the time domain spanning sinusoids of the Fourier transform \citep{addison2017}.

In all of our wavelet transforms, the basis wavelet is the Morlet wavelet, the product of a complex sinusoid with a Gaussian window. It may be written as \citep{moca2021}:
\begin{equation}
\Uppsi_{f,c}(t) = e^{-\frac{t^2}{2 B_c^2}} e^{2\pi i f t} 
\label{eq:morlet}
\end{equation}
where $B_c$ is the time spread parameter, which is proportional to the number of cycles $c$ and inversely proportional to the central frequency $f$:
\begin{equation}
B_c=\frac{c}{k_{sd}f} 
\label{eq:time_spread_param}
\end{equation}
In this paper we take $k_{sd} = 5$, the value of this parameter is simply a `design choice' in the Morlet wavelet. Given some time series $s(t)$, the continuous wavelet transform (CWT) may be calculated analogously to the Fourier transform, by performing a convolution of $s(t)$ with the complex conjugate of the wavelet $\Uppsi^*$:
\begin{equation}
\centering
T(a, b) = w(a)\int_{-\infty}^{\infty} s(t) \,\Uppsi^* \left( \frac{t - b}{a} \right) \,dt
\label{equ:wavelet_transform}
\end{equation}
where $w(a)$ is the weighting function which we take to be $(B_c\sqrt{2\pi})^{-1}$ as this normalisation ensures that the estimation of instantaneous power is independent of frequency \citep{moca2021}. The parameters $a$ and $b$ control the scale (dilation) and location (translation) of the wavelet, respectively. In a more practical sense, since $a$ determines the central frequency of the wavelet and $b$ controls the position in the time domain, these parameters may be substituted for the frequency $f$ and time $t$ to obtain the transform in time-frequency space $T(t, f)$. The squared modulus of $T(t, f)$ is the real valued wavelet scalogram, analogous to the Fourier periodogram. The wavelet scalogram can be used to identify and study transient signals in time series data, such as estimating their frequency and duration. The wavelet transform, like the Fourier transform, requires equally spaced data. It cannot be applied to data that is unevenly sampled.\footnote{If the data is unevenly sampled, as in \textit{Suzaku} light curves, then the WWZ transform \citep{foster1996} may be used.}

In the wavelet transform, the frequency resolution decreases at the higher frequencies, and likewise the temporal resolution decreases at lower frequencies. Increasing the number of cycles in the basis wavelet will improve frequency resolution and the cost of temporal resolution, and vice versa. This is an effect of the Heisenberg-Gabor uncertainty principle which states that the exact time and frequency of a signal cannot be measured \citep{gabor1946}.

Neither the short-time Fourier Transform (STFT) or CWT are at the limit of this principle. The time/frequency resolution of both the STFT and CWT may be improved using ``super-resolution'' techniques. Multiple spectrograms/scalograms calculated with good time and good frequency resolution may be combined into one \citep[e.g.][]{shafi2009, nam2010}. For the Fourier transform, the geometric mean of multiple spectrograms gives the minimum mean cross-entropy (MMCE) \citep{loughlin1994}. For the wavelet transform, the geometric mean of multiple scalograms gives the superlet transform (SLT) \citep{moca2021}.

A superlet is defined as a set of wavelets with a given central frequency $f$, spanning some number of different cycles to give a range of temporal and frequency resolutions:
\begin{equation}
\centering
S_{f,\,o} = \left\{ \Uppsi_{f,\,c}\,|\,c=c_1,c_2,...,c_o \right\}
\label{equ:superlet}
\end{equation}
where $o$ is the order of the superlet, determining the maximum number of cycles in the superlet. The order, and minimum cycles $c_1$ are to be chosen before calculating the SLT. Note that an SLT with order $o=1$ is equivalent to the CWT with $c_1$ cycles.

The superlet response to a signal $s(t)$ is the geometric mean of the responses to each individual wavelet in the superlet set:
\begin{equation}
\centering
R\left [S_{f,\,o}\right ] = \left(\prod^o_{i=1}R\left [\Uppsi_{f,\,c_i}\right ]\right)^{1/o}
\label{equ:superlet_response}
\end{equation}

The response to each wavelet is the complex convolution of the wavelet with the signal (analogous to Eq. \ref{equ:wavelet_transform}):
\begin{equation}
\centering
R\left [\Uppsi_{f,\,c_i}\right ] = \sqrt{2}\,s(t)\,*\,\Uppsi_{f,\,c_i}
\label{equ:superlet_conv}
\end{equation}
As in the wavelet transform, the squared modulus of the superlet response gives the real valued superlet scalogram, which can be used and interpreted in the same way as the wavelet scalogram, by plotting response power against frequency and time.

For the application of the superlet transform to astrophysics, and in particular X-ray light curves, we introduce \textsc{XSuperlet}\footnote{\href{https://github.com/THoddAstro/XSuperlet}{github.com/THoddAstro/XSuperlet}} \citep{hodd2026}, a Python program for calculating the wavelet and superlet transforms of light curves. Unless otherwise stated, in this paper we used superlets with minimum cycles $c_1 = 3$ and order of $o\mathrm{:}2-6$. Fewer minimum cycles leads to single continuous signals appearing as multiple shorter signals in the scalogram, whilst a higher order biases the transform to higher frequency resolution at the cost of some temporal resolution.

The cone of influence (COI) describes the region wherein the wavelet or superlet transform is able to sufficiently constrain its values. Outside of the COI, scalogram values are subject to boundary effects and may not be a valid representation of the input time series. This occurs due to the length of the input data, with lower frequencies requiring more time to constrain properly. The result is that low frequency signals can only be studied towards the centre of the time series. Since there is no exact definition to calculate the COI, we adopt the expression used by \cite{torrence1998}:
\begin{equation}
\centering
\mathrm{COI} = \frac{6\sqrt{2}}{2\pi f}
\label{equ:coi}
\end{equation}
On all our scalograms, the regions outside the COI are shaded in dark grey to show where caution should be applied to any interpretation of the data.

\begin{figure}
	\includegraphics[width=\columnwidth]{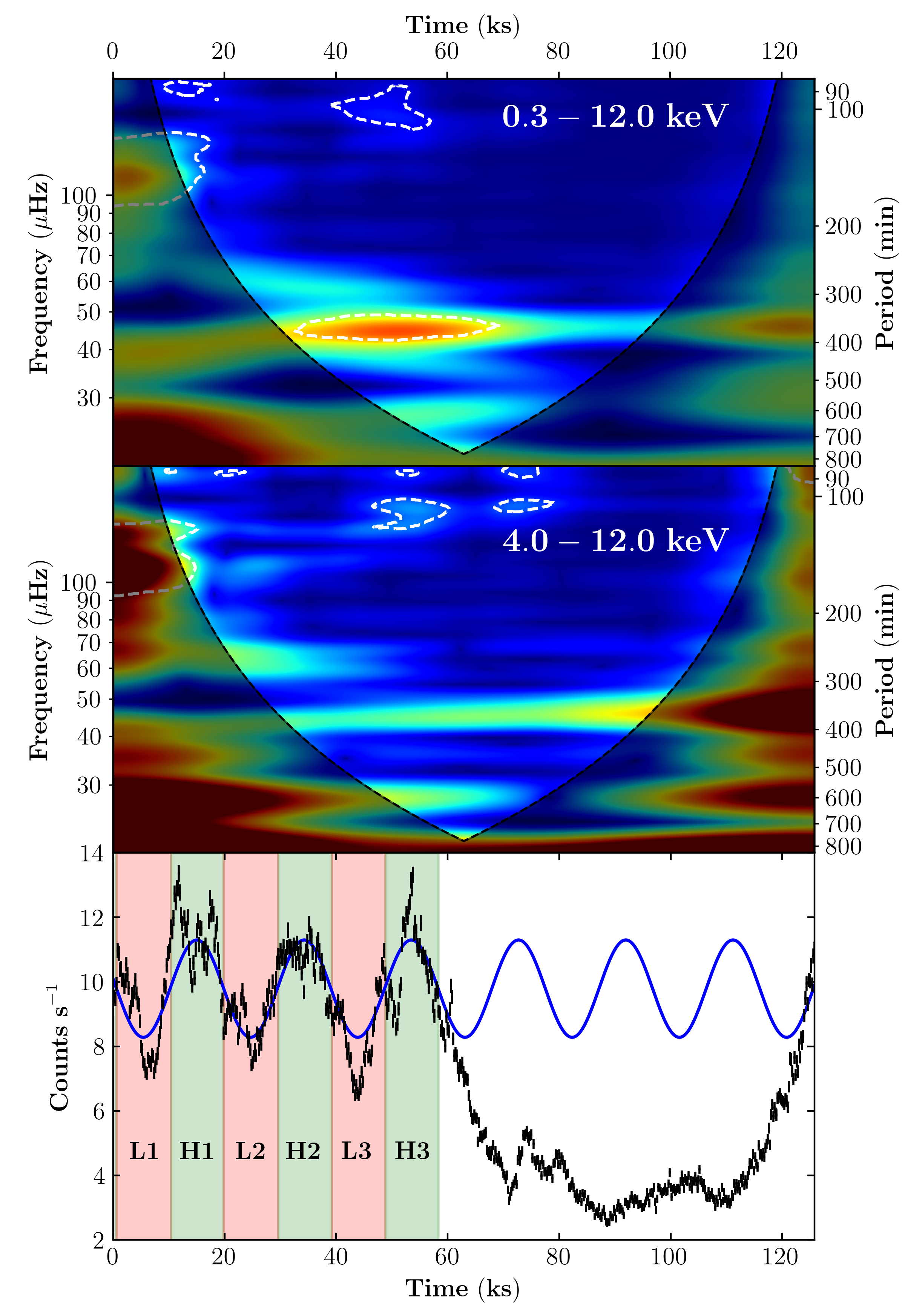}
    \caption{Top: The SLT for the broad band $0.3-12.0$\,keV light curve. Dashed contours show the 90\% confidence regions of the detected signal. The dark shaded regions denote areas outside the COI. The colour corresponds to the measured wavelet power, from low (blue) to high (red), and the units on wavelet power are arbitrary and scaled to best show the features of each scalogram. Middle: The SLT for the hard band $4.0-12.0$\,keV light curve. Bottom: A sine curve fitted to the first 60\,ks of the $0.3-12.0$\,keV light curve and and then propagated to the end of the observation. In Section~\ref{sec:phase_spec}, phase-resolved spectra are examined for the average high (green) and average low (red) phase. In Section~\ref{sec:6phase_spec}, phase-resolved spectra are examined for each peak/trough interval (i.e. six spectra from L1-H3).}
    \label{fig:slt_stack}
\end{figure}

To quantify the periodic behaviour in the light curve of NGC~6814, we applied the SLT to each of the evenly binned light curves using \textsc{XSuperlet}. Fig.~\ref{fig:slt_stack} shows the resulting scalograms. The most significant feature in the broad band SLT is a strong signal between the COI at 30\,ks and around 70\,ks at approximately $40-50$\muhz. This feature is also seen in the soft band ($0.3-1.0$\,keV) SLT (Fig.~\ref{fig:slt_soft_p31}) and to a lesser extent in the 2021 \textit{XMM-Newton} and 2011 \textit{Suzaku} observations (Figs. \ref{fig:slt_21}, \ref{fig:wwa_11}). We estimate this frequency by taking the full width half maximum (FWHM) of the instantaneous power spectrum at $50$\,ks. This yields a frequency of $45\err{6}{4}$\muhz, corresponding to a timescale of $22\err{3}{2}$\,ks ($320-400$ minutes) for the periodic behaviour.

The significance of signal detection was estimated by simulating 1000 light curves using the methods of \cite{timmer1995} and \cite{emmanoulopoulos2013}. These light curves were generated by first fitting the power spectrum (PSD) and probability distribution function (PDF) of the real light curve. The PSDs were calculated by taking the square of the DFT of the light curve, and fit using a smoothly bending power law. The PDFs were fit using kernel density estimation (KDE) with a Gaussian kernel and \textsc{scikit-learn} \citep{scikit-learn}. Simulated light curves are then created by sampling light curves with the same duration, time bins, PSD, and PDF of the real light curve. The SLT was then calculated for each simulated light curve using the same parameters, and we determined where the scalogram of the real light curve had significantly greater power than the $n^{\mathrm{th}}$ percentile of simulated scalograms. Using this method, the detection of periodic behaviour is only \textgreater$90\%$ significant prior to the eclipse at $\sim60$\,ks. We find that the detection in the soft band is more significant than the broad band.

Since the eclipse is strongest in the soft band, the reduction in wavelet power during the eclipse is expected. To see the effect of the eclipse in the hard band, we calculated the SLT of the hard band light curve (Fig.~\ref{fig:slt_stack}). Here we see that the periodic signal is not reduced by the eclipse, and in fact continues for the entire duration of the eclipsing event, even increasing towards the end of the observation. We note that none of the hard band SLT was found to be statistically significant among 1000 other simulated light curves, to the 90\% level, and the wavelet power is much lower.

\begin{figure}
	\includegraphics[width=\columnwidth]{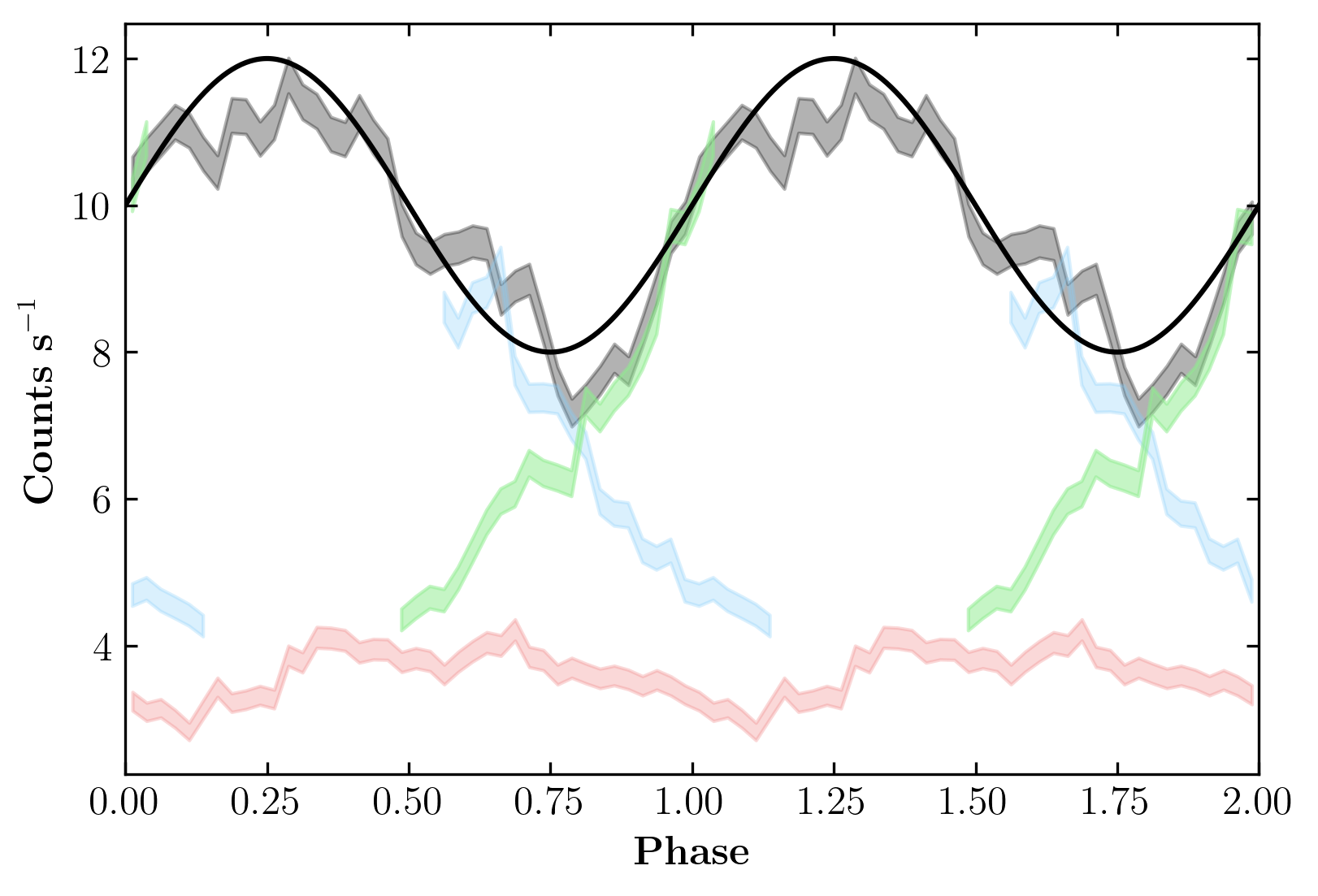}
    \caption{The broad band phase-folded light curve. The light curve is folded with a period of $19$\,ks (52\muhz). The black band shows data prior to the eclipse ingress. Blue, red, and green bands show data during the ingress, eclipse, and egress, respectively. The shaded areas are the regions of uncertainty on the count rate of each bin. The black curve is a sinusoid with the same period and phase as the folded data. Two periods are included for clarity.}
    \label{fig:phasefold}
\end{figure}

Given the possible oscillation identified by the SLT, we next fit a sinusoid to the broad-band light curve with initial parameters from the SLT detection. We employed a simple sinusoid model and chi-squared minimisation. In this fit (Fig.~\ref{fig:slt_stack}) we see that the three peaks and troughs at the start of the observation are well matched, and the egress of the eclipse is also fit. This fit gives a frequency of $52\pm{1}$\muhz, which is consistent with that found by the SLT. We also plot the light curve phase-folded to a $52$\muhz\ frequency (Fig.~\ref{fig:phasefold}). Note that at this frequency, the ingress and egress of the eclipse are in-phase with the oscillation. The period identified also corresponds to significant peaks in the broad band and soft band PSDs (Fig.~\ref{fig:psd_broad}).

The low-power feature between 20 and 40\,ks at $60-70$\muhz\ is seen in all scalograms, but is only \textgreater90\% significant in the mid band (Fig.~\ref{fig:slt_mid}). It is possible that interference from this feature is responsible for the frequency of the best-fit sinusoids being slightly above that found by the SLT pre-eclipse.

\subsection{Lag-Frequency Analysis}
\label{sec:lag}
Having identified the periodic behaviour in NGC~6814, we next used the X-ray timing package \textsc{pyLag}\footnote{\href{https://github.com/wilkinsdr/pyLag}{github.com/wilkinsdr/pyLag}} to search for any reverberation lags that may help us understand the origin of this oscillation. The Fourier lag-frequency and lag-energy spectra may be used to identify lags between energy bands at given frequencies \citep{uttley2014}. Reverberation lags occur as a result of the light travel times between the primary emission from the hot corona and the reflected emission from the disc and warm corona. In principle, we would expect intrinsic variations in the hot corona to lead the same variations in the reflected emission from the disc due to the shorter path travelled by the primary emission. The warm corona may also produce some intrinsic variability in the soft band.

For this section we will employ the discrete Fourier transform (DFT) of our evenly binned light curves and the method outlined by \cite{uttley2014}. The DFT of light curve $x(t)$ may be written as:
\begin{equation}
X_n = \sum^{N-1}_{k=0}x_k\,e^{\frac{2\pi ink}{N}}
\label{equ:dft}
\end{equation}
$X_n$ is the value of the DFT at $f_n=n/(N\Delta t)$ where the minimum frequency is $f_{\textrm{min}}=1/(N\Delta t)$ (the inverse of observation duration), and the maximum is $f_{\textrm{max}}=1/(2\Delta t)$ (the Nyquist frequency). The normalised periodogram is:
\begin{equation}
P_n = \frac{2\Delta t}{\langle x\rangle^2N} X_n^*X_n
\label{equ:periodo}
\end{equation}
To find the frequency dependent lag between two light curves ($x$ and $y$) we can calculate the Fourier cross-spectrum, which we may write in complex polar form:
\begin{equation}
C_{XY,n} = X_n^*Y_n = A_{X,n}A_{Y,n}\,e^{i\phi_n}
\label{equ:cross_spectrum}
\end{equation}
Here, $A_{X,n}$ and $A_{Y,n}$ are the moduli of $X_n$ and $Y_n$, whilst $\phi_n\equiv \textrm{arg}(C_{XY,n})$ is the phase lag between the two light curves from which we can determine the time lag within each frequency bin to produce the Fourier lag-frequency spectrum. 

\begin{figure}
	\includegraphics[width=\columnwidth]{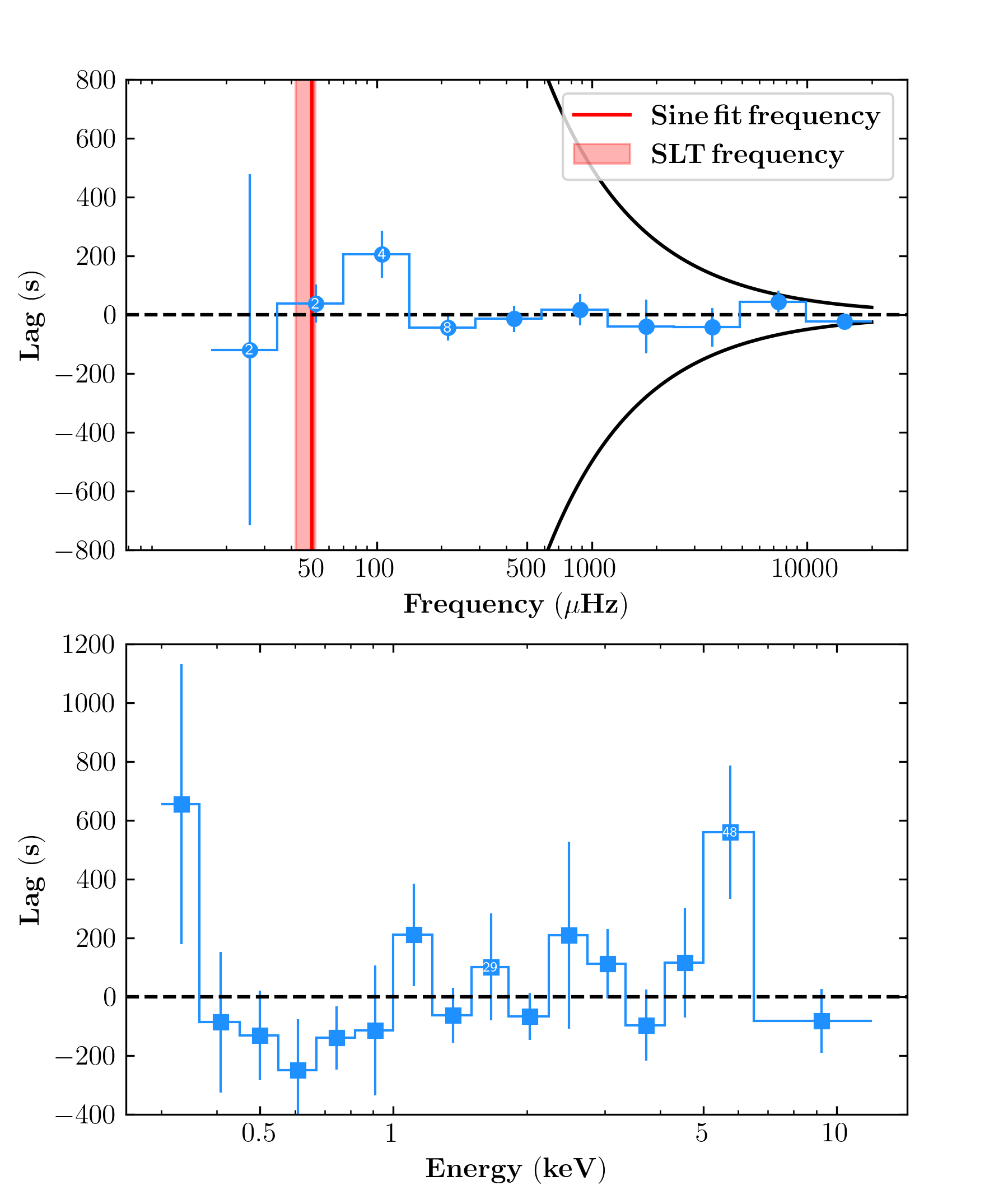}
    \caption{Top: The lag-frequency spectrum between the $0.3-1.0$\,keV and $1.0-4.0$\,keV light curves prior to the eclipse. The reported period is plotted as a red band. The digits in the data points correspond to the number of measurements in each bin for those with less than ten measurements. The black curve denotes the region outside of which the effects of phase wrapping should be considered. Bottom: Lag-energy spectrum surrounding $50$\muhz, the frequency of the identified periodic behaviour, for the first $60$\,ks of the observation. The digits in the data points correspond to the number of measurements in each bin for those with less than 100 measurements.}
    \label{fig:lag_analysis}
\end{figure}

For our NGC~6814 light curves, we used 10 evenly spaced frequency bins over the range $17-20000$\muhz\;and calculated the lags between the $0.3-1.0$\,keV and $1.0-4.0$\,keV light curves. The results of this are shown in the top panel of Fig.~\ref{fig:lag_analysis}. There is only one indication of a possible hard band lag of $\sim200$\,s (by convention, a positive lag indicates the hard band lags behind the soft band) at $\sim100$\,\muhz. However, this signal contains only 4 data points, therefore its importance is dubious.

We next calculated the Fourier lag-energy spectrum for the pre-eclipse part of the observation by cutting the light curves down to the first 60\,ks only. We used our 16 narrow energy band light curves (see Section \ref{sec:obs}) and calculated the lag between them and the broad $0.3-12.0$\,keV reference band at a frequency of $40-70$\,\muhz. This gives some indication that in the hard bands $\sim5-6$\,keV oscillations of $\sim50$\,\muhz\ may lag behind the reference band (Fig.~\ref{fig:lag_analysis}, bottom), although the significance of this detection is low ($<2\sigma$). 

The lag-energy spectrum was also calculated during the eclipse and for the entire light curve.  No delays were detected in either case. To check for the effects of binning, we calculated lag-frequency and lag-energy spectra with alternative binning parameters (e.g. fewer/more bins, different frequency bounds). In all cases there were no significant changes or lags detected.

\subsection{Discrete Correlation Functions}
\label{sec:dcf}
We examine for frequency-independent lags between between energy bands using the discrete correlation function (DCF) \citep{edelson1988}. Adopting a slightly modified version of \texttt{pyDCF}\footnote{\href{https://github.com/astronomerdamo/pydcf}{github.com/astronomerdamo/pydcf}} \citep{robertson2015}, the DCF between the soft ($0.3-1.0$\,keV) and very hard ($6.0-12.0$\,keV) bands was calculated for both pre-eclipse and eclipse light curves.  To estimate the significance of the correlations, $10^6$  pairs of light curves (1000 soft band and 1000 hard band) were simulated as described above. We then found the 68\%, 95\%, and 99.7\% percentile correlations in each time bin to identify the $1\,\sigma$, $2\,\sigma$, and $3\,\sigma$ significance levels for both the pre-eclipse and eclipse DCFs.

\begin{figure}
	\includegraphics[width=\columnwidth]{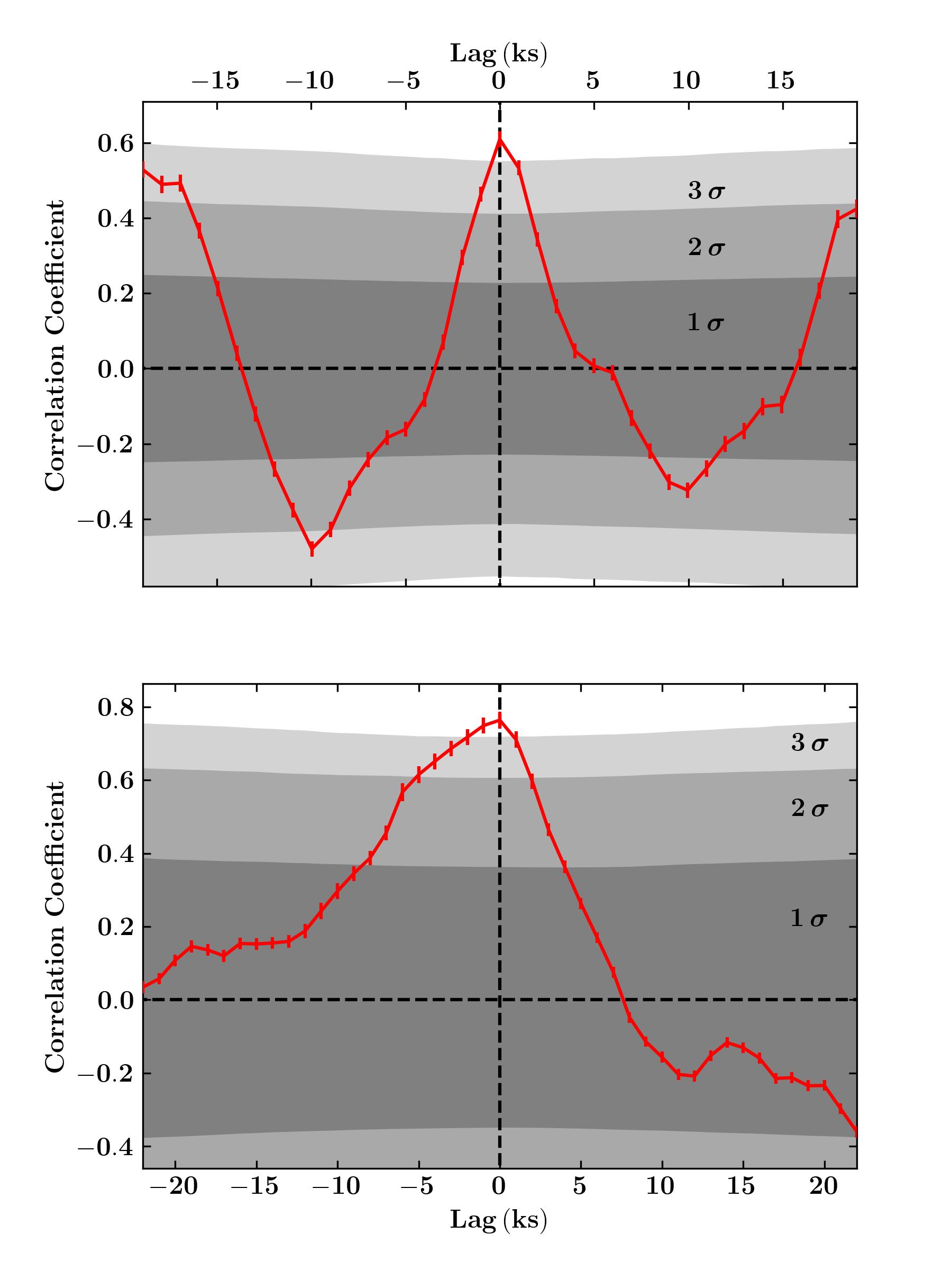}
    \caption{Top: The discrete correlation function of the soft and hard band light curves prior to the eclipse. The correlation peaks at zero seconds with 99.7\% ($3\,\sigma$) significance. Additional, strong peaks at $\pm19$\,ks confirm the oscillation period found above. Bottom: The discrete correlation function of the soft and hard band light curves during the eclipse. The correlation still peaks at zero seconds (with 99.7\% significance), but there is an asymmetry in the peak toward negative lags indicating the hard band leads the soft.}
    \label{fig:dcf_stack}
\end{figure}

In the first 60\,ks (Fig.~\ref{fig:dcf_stack}, top), we see a significant ($3\,\sigma$) positive correlation at zero seconds, and additional peaks at $\sim \pm19$\,ks.  This is commensurate with the periodic behaviour. However, during the eclipse (Fig.~\ref{fig:dcf_stack}, bottom), the correlation at zero seconds is still strong, but there is a significant asymmetry in the DCF toward negative delays.  Such a delay can be understood if some of the soft photons arrive at the observer after the hard photons.

\section{Modulation of the period in a dense cloud during the eclipse}
\label{sec:mod}
Inspection of the wavelet transform in Fig.~\ref{fig:slt_stack} shows that in the soft band the period is stronger and more significant prior to the eclipse.  Though the signal-to-noise is limited in the hard band ($4.0-12.0$\,keV) where the errors are relatively large compared to the count rate, the period appears to be more persistent and perhaps even stronger during the eclipse. Recall, that the column density of the eclipsing medium is sufficient to absorb X-rays below approximately 6\,keV \citep[]{gallo2021} so it may be interesting to examine if the period is impacted by the eclipse.

\begin{figure}
	\includegraphics[width=\columnwidth]{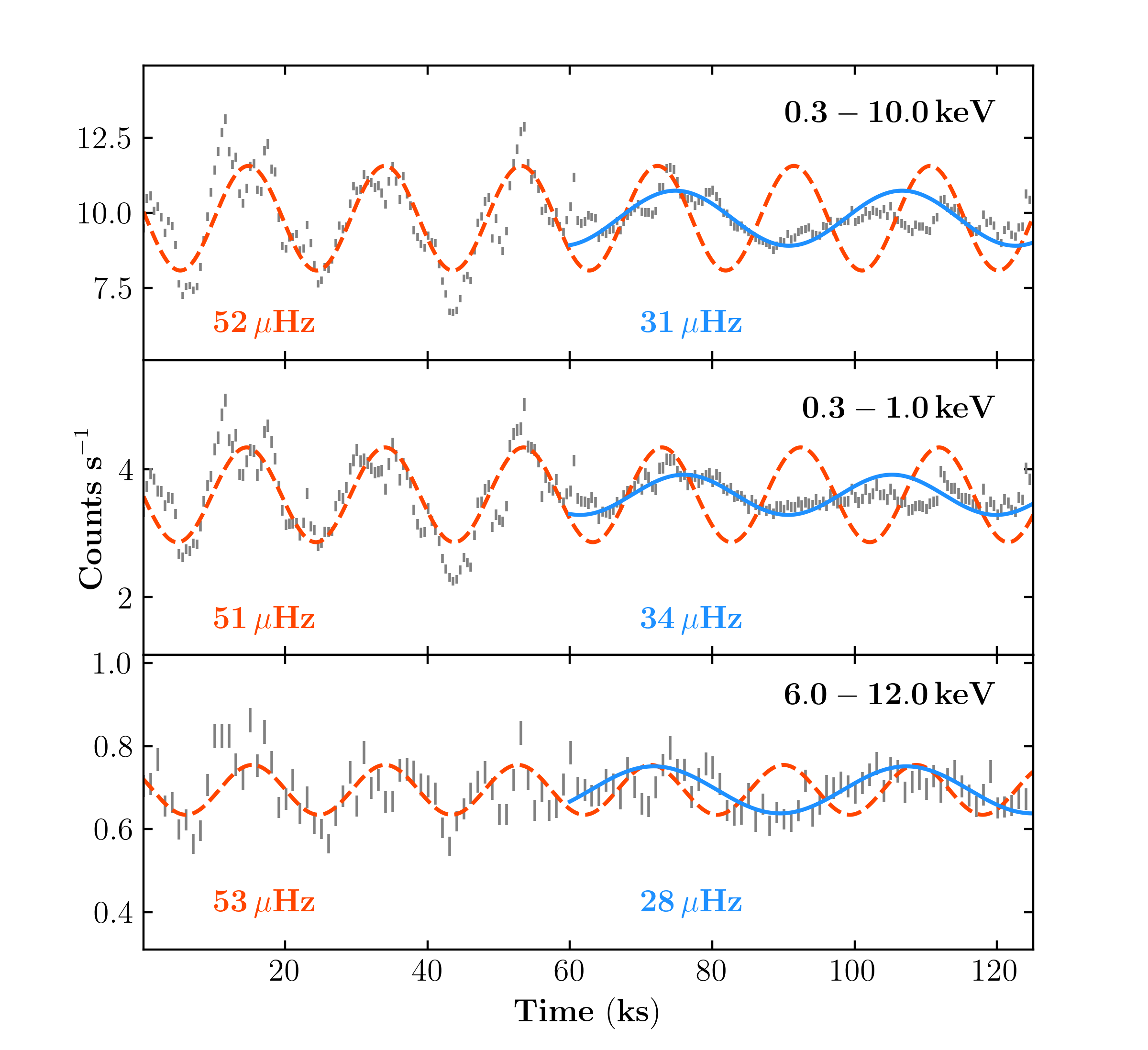}
   \caption{The normalised (eclipse removed) light curves in the broad ($0.3-10.0$\,keV), soft ($0.3-1.0$\,keV), and very hard ($6.0-12.0$\,keV) bands. The orange curve denotes the frequency found by fitting a sinusoid to the first 60\,ks of each light curve. These frequencies are $52\pm1,\ 51\pm1,\ 53\pm{1}$\muhz\ respectively, with errors estimated using Markov chain Monte Carlo (MCMC) methods. The blue curve is the sinusoid fit to the eclipse (60\,ks onwards) only. These frequencies are $31\pm{1},\ 34\pm{2},\ 28\pm{1}$\muhz\ respectively, showing that the modulation of the frequency during the eclipse occurs in all energy bands.}
   \label{fig:normlc}
\end{figure}

In Fig.~\ref{fig:normlc}, the (broad band) light curve is normalised to remove the effects of the eclipse. This was done by assuming a linear model for ingress and egress, and adjusting the counts of the eclipse to match the mean count rate during the first 60\,ks. Based on the superlet detection of a period with 90\% significance, we next fit a sinusoidal model to the first 60\,ks of each normalised light curve. This yields good fits at all energies. However, extrapolating the model to the end reveals a stretching of the data during the eclipse. This is confirmed by separately fitting a sinusoid to the latter 60\,ks only, which shows that in the second half of the observation the best fit sinusoid has a lower frequency ($\sim30$\,\muhz) compared to $\sim50$\,\muhz\ during in the first 60\,ks. This behaviour appears in all the energy bands (Fig.~\ref{fig:normlc}). We note that this apparent change in frequency may be non-physical and instead due to red noise fluctuations \citep{vaughan2016}.

The apparent decrease in frequency is also visible in some of the superlet scalograms as a second, low power, signal (see Appendix \ref{sec:slt_appendix}) at frequencies of around $30$\,\muhz\ though it is never significant. Some of the scalograms suggest such a signal may exist prior to the eclipse, though at that time the $\sim50$\,\muhz\ signal dominates, with the lower frequency signal outside of or close to the COI.

We used the time-dependent photoionisation model (\textsc{tpho}) \citep{rogantini2022} to investigate whether the finite response time of the obscurer to a periodic ionising luminosity could alter, and in particular broaden, the observed periodic signal. Unlike most ionised absorber models which assume equilibrium between the flux and the ionisation of the absorber, \textsc{tpho} is a non-equilibrium model. In a photoionised plasma, the ionisation state does not necessarily respond instantaneously to variations in the incident flux. The response depends strongly on the gas density: high-density gas adjusts rapidly because its recombination timescales are short, while low-density gas responds more slowly and could in principle introduce delays or smooth out the variability signal.

To test this, we simulated light curves produced by a periodic source seen through an obscuring wind with \textsc{tpho}. We explored covering factors between 0.5 and 1.0, column densities between 2 and $17.5\times 10^{22}\,\text{cm}^{-2}$, and gas densities between $10^{6} - 10^{10}\,\text{cm}^{-3}$. No significant distortion of the input periodic signal is found in any of the tested cases. This suggests that time-dependent photoionisation effects are too small to significantly affect the observed light curve.

\section{Spectral Analysis}
\label{sec:spectral}
According to the sine curve fit in Figure~\ref{fig:slt_stack}, we phase-resolved the first 60\,ks of the observation into three peaks (H1-H3) and three troughs (L1-L3) based on their chronological appearance in the light curve. 

\begin{figure}
	\includegraphics[width=\columnwidth]{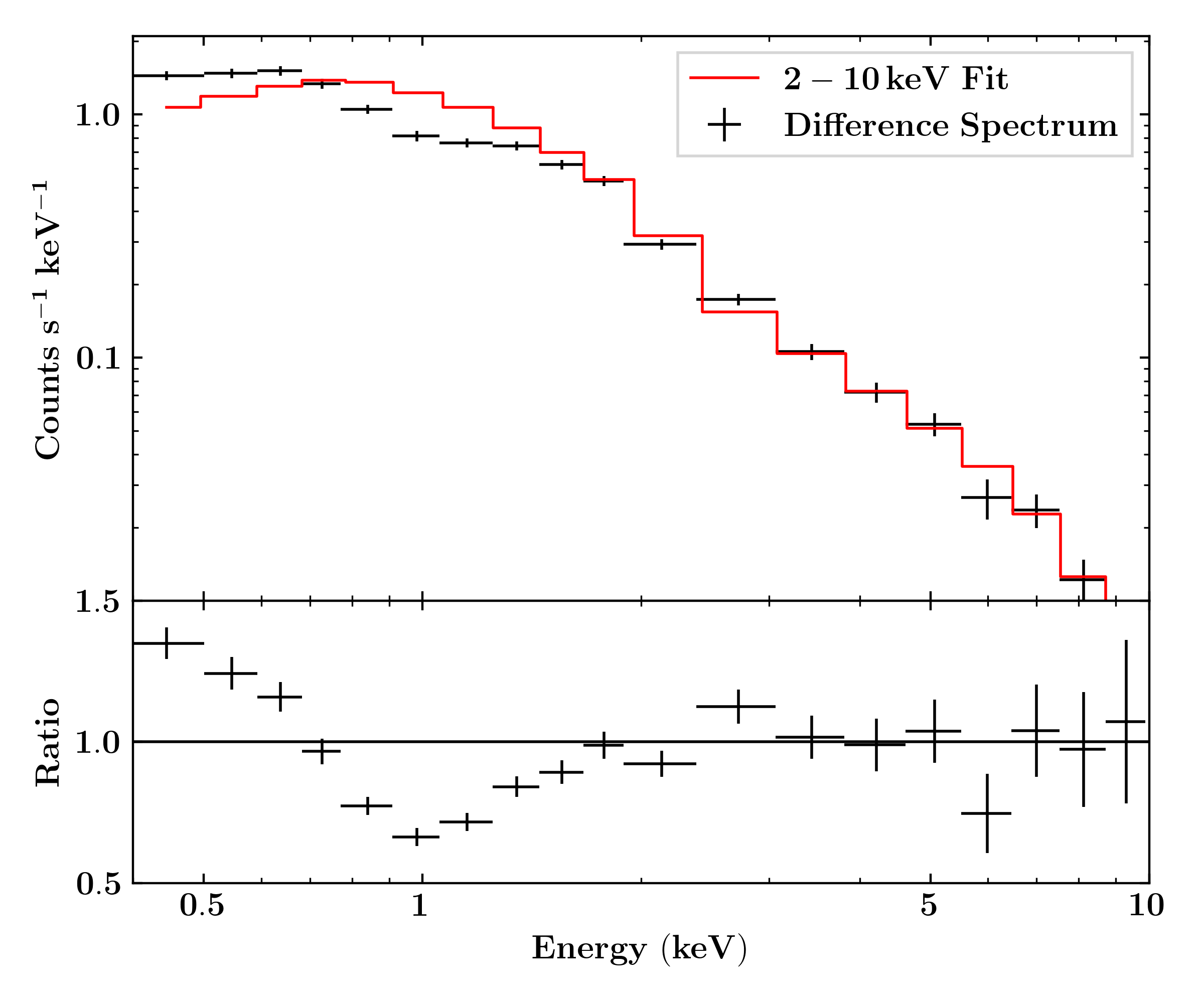}
    \caption{The high-phase spectrum subtracted by the low-phase spectrum (i.e. difference spectrum). A simple \texttt{tbabs\,\texttimes\,po} model is fit to the $2-10$\,keV data, and extrapolated to lower energies revealing significant deviations below $1$\,keV.}
    \label{fig:diffspec}
\end{figure}

To verify the spectral differences between the peaks and the troughs, we combined the peaks into a single `high phase' spectrum, and the troughs into a `low phase' spectrum, and calculated the difference of the two (Fig.~\ref{fig:diffspec}). This difference spectrum was fitted in the $2-10$\,keV band with a power law ($\Gamma\approx1.97$) absorbed by Galactic column density and then the fit was extrapolated to lower energies to highlight regions where the spectral changes occur. The spectrum deviates from a power law most strongly at $\sim 1$\,keV and $\sim 0.5$\,keV. This could be indications of changes in absorption, which will be examined for in the spectral models. We will start by considering the combined high phase and low phase spectra to see the general differences between the peaks and troughs of the oscillation. Then we will investigate all six phases individually to understand if there are any differences between each high/low phase.

\subsection{Average Phase Resolved Spectra}
\label{sec:phase_spec}
We used \textsc{xspec} to apply the Comptonisation model from \cite{gallo2021}. This model attributes the soft excess to a warm corona that sits some height above the inner accretion disc. In particular, this model consists of two power laws (the hot and warm coron\ae) both modelled by \textsc{nthcomp} components. The hot corona has a shallower power law slope and is optically thin whilst the warm corona is steeper and optically thick. Additionally, two warm absorbers (XABS1 \& XABS2) are included based on RGS spectral modelling \citep{gallo2021}. These are added using the \textsc{xspec} implementation \citep{parker2020} of the \textsc{spex} \citep{kaastra1996} model \textsc{xabs} \citep{steenbrugge2003}. We use the large \textsc{xabs} table model and each \textsc{xabs} component is fully covering. \textsc{zxipcf} is included as a partially ionised partial coverer, and we also model a Gaussian profile at $\sim6.45$\,keV.

We begin by modelling the combined high phase and low phase spectra. Phases were identified using the sinusoid fit to the light curve in Section~\ref{sec:timing}. Allowing only the normalisations of the power law components to change establishes a base model where $C=502$ for $228$ dof. Allowing the covering fraction of the partial coverer to change between the high and low phases, improves the fit significantly with $C=270$ with $226$ dof and  $\Delta\mathrm{AIC}=-228$. Alternatively, describing the difference between the two phases with free power law normalisations and free hot corona photon index results in an equally good statistical fit with $C=273$ with $226$ dof and $\Delta\mathrm{AIC}=-225$. 

Consequently, we have two possible models that describe the differences between the low and high phases (see Table~\ref{tab:model12}): (i) Changing covering fraction (higher covering fraction during troughs), and (ii) Changing hot corona power law (shallower during troughs). Both models have similar statistical quality.

\subsection{Six Phase Resolved Spectra}
\label{sec:6phase_spec}
The two possible models that could describe the average high-phase and low-phase spectra (Section~\ref{sec:phase_spec}), are now tested to determine if they can describe the time variability. The time scale of importance is $\sim10$~ks as the source moves from low-phase to high-phase. The six spectra derived from the three peaks (H1-H3) and three troughs (L1-L3) labelled in Figure~\ref{fig:slt_stack}, are fitted with each of the two models described in Section~\ref{sec:phase_spec}. Table~\ref{tab:model12} lists the measured parameters that are not varying between the phase-resolved spectra. The parameters that are varying over the changes in phase are marked with an asterisk in Table~\ref{tab:model12} and reported in Table~\ref {tab:freepar12}.

\begin{figure}
	\includegraphics[width=\columnwidth]{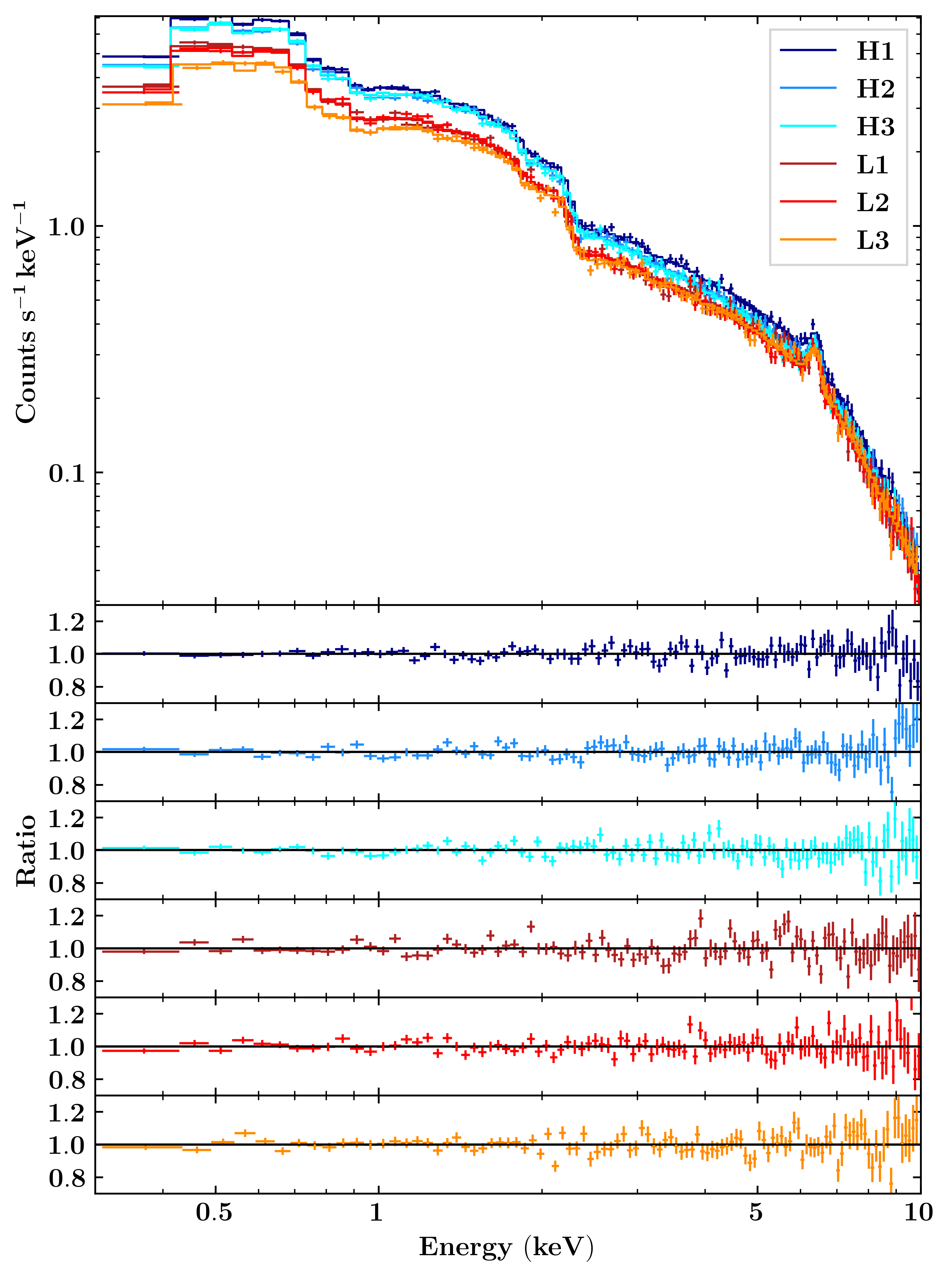}
    \caption{The six phase-resolved spectra fit to Model 1 (i.e. changes in covering fraction).}
    \label{fig:6spec}
\end{figure}

In Model~1, the parameters that were permitted to vary over the six phase changes were the covering fraction of the ionised partial coverer and the normalisations on the two \textsc{nthcomp} components.  This resulted in a fit statistic of $C=649$ for 622 dof. The fit and residuals of each phase fit to Model 1 is given in Figure \ref{fig:6spec}.

Alternatively, Model~2 was also applied to the six phase resolved spectra.  In this case, the photon index of the hot corona and two power law normalisations were permitted to vary. The fit yielded  a similar statistic as Model~1 with $C=656$ for 621 dof ($\Delta\mathrm{AIC}=+8$). Other permutations of the models, where different parameters were allowed to vary, were also tested. However, the best fits obtained were with the variations of the Model~1 and 2 described above.

\begin{figure}
	\includegraphics[width=\columnwidth]{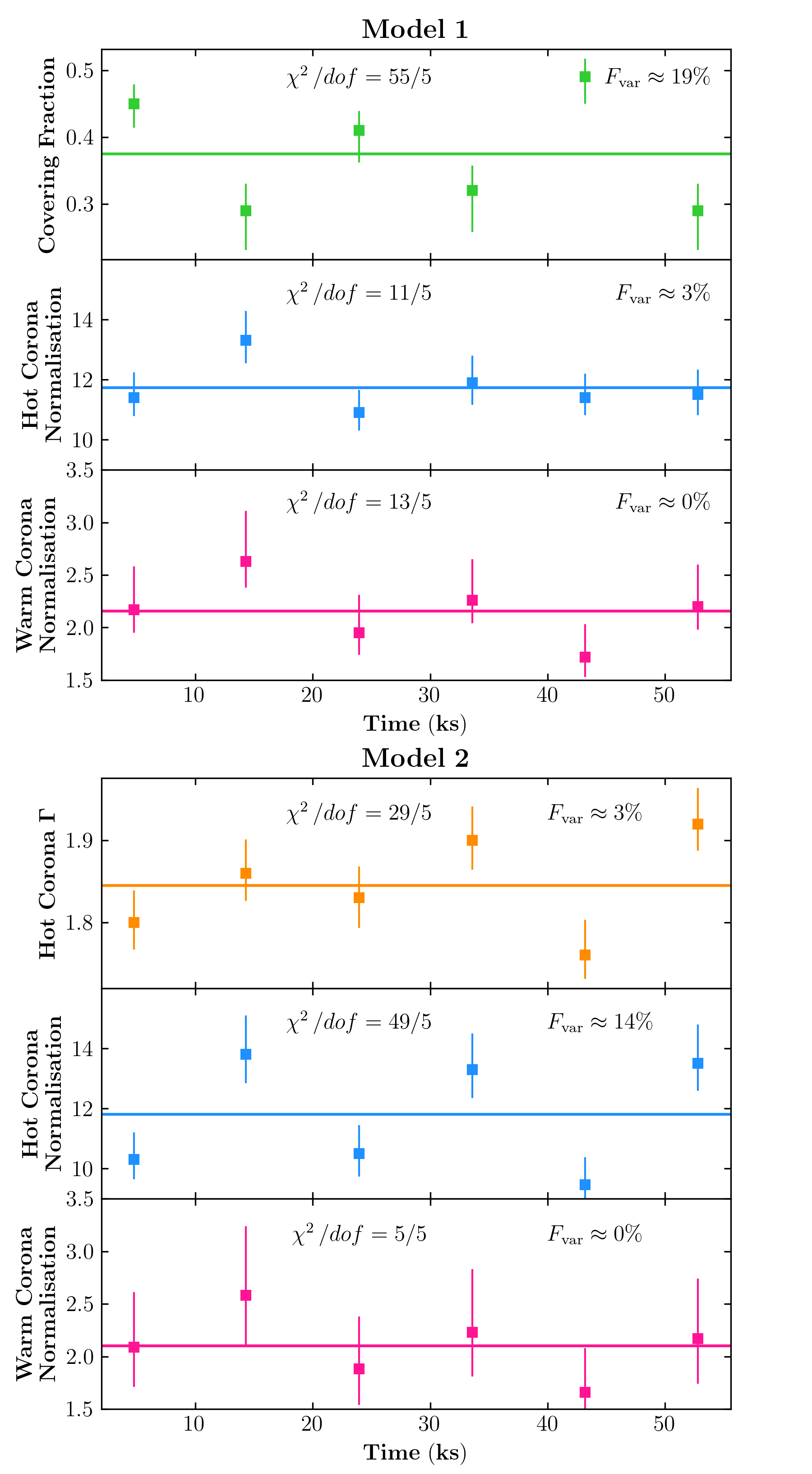}
    \caption{Top: Model 1 best fit parameter values for covering fraction, warm corona normalisation, and hot corona normalisation. Bottom: Model 2 best fit parameter values for hot corona power law index, warm corona normalisation, and hot corona normalisation. Normalisations are in units of $10^{-3}$. The $\chi^2/dof$ are calculated assuming a constant fit to the mean parameter value (horizontal lines). In each panel, the corresponding $F_{\text{var}}$ for that parameter is shown.} 
    \label{fig:modpars}
\end{figure}

In Figure~\ref {fig:modpars}, the varying parameters in Model 1 and 2 are plotted as a function of time. For each parameter we calculate the fractional variability \citep{vaughan2003} to estimate the amplitude of variability after accounting for the measurement uncertainties. In Model~1, the most variable parameter is the covering fraction ($F_{\text{var}}\approx19\%$). As in the average phase-resolved spectral analysis (Section~\ref{sec:phase_spec}), the covering fraction of the ionised absorber fluctuates from about $0.3$ in the high-phase (when brighter) to about $0.45$ in the low-phase (when dimmer).  In Model~2, the most variable component is the normalisation of the hot corona ($F_{\text{var}}\approx14\%$), which is higher in the bright high-phase, and lower in the dim low-phase.  The other parameters vary to a lesser degree, but appear correlated with hot corona normalisation changes.  That is, as the hot corona normalisation increases, so does its photon index and the normalisation of the warm corona. This is consistent with the softer-when-brighter effect \citep[e.g.][]{sobolewska2009, serafinelli2017}. Despite being consistent with a fixed value in Model~2, linking the warm corona normalisation across all states yields a significantly worse fit of $C=781$ for 626 dof ($\Delta\mathrm{AIC}=+115$). Doing the same in Model~1 also results in a poorer fit of $C=794$ for 627 dof ($\Delta\mathrm{AIC}=+135$).

\begin{table*}
\renewcommand{\arraystretch}{1.25}
    \centering
    \caption{Partial covering Comptonisation model applied to each phase. \textsc{xabs$_1$} and \textsc{xabs$_2$} are two warm absorbers \citep{gallo2021}, while \textsc{nthcomp$_{\mathrm{soft}}$} and \textsc{nthcomp$_{\mathrm{hard}}$} are the two power law components from the warm and hot coron\ae\space, respectively. In \textsc{xspec} terminology the model is \texttt{tbabs\,\texttimes\,ztbabs\,\texttimes\,xabs$_1$\,\texttimes\,xabs$_2$\,\texttimes\,(zxipcf\,\texttimes\,(nthcomp$_{\mathrm{soft}}$\,+\,nthcomp$_{\mathrm{hard}}$)\,+\,zgauss)}. Parameters that differ in the high- and low-phase are reported in the same row as ``high / low''. }
    \begin{tabular}{llcc}
        \hline
        \textbf{Model }                     & \textbf{Model}                                               & \textbf{Model 1} & \textbf{Model 2}\\
        \textbf{Component}                  & \textbf{Parameter}                                            & \textbf{Parameter Value}  & \textbf{Parameter Value}\\ \hline \hline
        \textsc{ztbabs}                     & $N_H/10^{20}\,\mathrm{cm}^{-2}$                               & $5.56\err{1.30}{0.93}$  & $4.92\err{2.00}{1.70}$  \\ \hline
        \textsc{xabs$_1$}                   & $\mathrm{log}\,\xi/\mathrm{erg\,cm}^{-2}\,\mathrm{s}^{-1}$    & $2.91\err{0.07}{0.06}$  & $2.88\err{0.07}{0.06}$ \\ 
                                            & $N_H/10^{21}\,\mathrm{cm}^{-2}$                               & $23.5\err{3.1}{5.8}$ & $23.5\err{3.9}{5.0}$  \\ 
                                            & $v_{out}/\textrm{km\,s}^{-1}$                                 & $9900\err{1920}{1680}$ & $13350\err{1710}{1830}$ \\ \hline
        \textsc{xabs$_2$}                   & $\mathrm{log}\,\xi/\mathrm{erg\,cm}^{-2}\,\mathrm{s}^{-1}$    & $1.10\err{0.09}{0.12}$  & $1.15\err{0.15}{0.08}$ \\ 
                                            & $N_H/10^{21}\,\mathrm{cm}^{-2}$                               & $4.22\err{0.28}{0.50}$ & $4.54\err{0.35}{0.37}$ \\ 
                                            & $v_{out}/\textrm{km\,s}^{-1}$                                 & $3690\err{1980}{2610}$ & $2955\err{2910}{1680}$ \\ \hline
        \textsc{zxipcf}                     & $N_H/10^{22}\,\mathrm{cm}^{-2}$                               & $14.1\err{1.0}{1.5}$ & $14.6\err{1.1}{1.7}$      \\ 
                                            & $\mathrm{log}\,\xi/\mathrm{erg\,cm}^{-2}\,\mathrm{s}^{-1}$    & $2.02\err{0.07}{0.06}$ & $2.06\pm{0.08}$ \\ 
                                            & $f_{cov}$                                                     & $0.300\err{0.010}{0.020}$ / $0.450\pm{0.040}$ & $0.387\err{0.041}{0.049}$  \\  \hline
        \textsc{nthcomp$_{\mathrm{soft}}$}  & $\Gamma$                                                      & $2.4$ & $2.02\err{0.38}{0.35}$  \\ 
                                            & $kT_e/\mathrm{eV}$                                            & $122\err{4}{9}$ & $124\err{9}{1}$  \\ 
                                            & $kT_{bb}/\mathrm{eV}$                                         & $3$ & $3$ \\ 
                                            & $\mathrm{Norm}/10^{-3}$                                       & $2.36\err{0.16}{0.27}$ / $1.95\err{0.23}{0.22}$ & $2.33\err{0.16}{0.25}$ / $1.88\err{0.22}{0.21}$ \\ \hline
        \textsc{nthcomp$_{\mathrm{hard}}$}  & $\Gamma$                                                      & $1.85\pm{0.03}$ & $1.89\pm{0.03}$ / $1.80\err{0.04}{0.03}$ \\
                                            & $kT_e/\mathrm{keV}$                                            & $100$ & $100$ \\ 
                                            & $kT_{bb}/\mathrm{eV}$                                         & $3$ & $3$ \\ 
                                            & $\mathrm{Norm}/10^{-3}$                                       & $12.2\err{0.7}{1.1}$ / $11.2\err{0.3}{0.2}$& $13.5\err{0.2}{0.3}$ / $10.1\err{0.6}{0.4}$ \\ \hline
        \textsc{zgauss}                     & $E/\mathrm{keV}$                                              & $6.42\pm{0.02}$ & $6.41\err{0.02}{0.01}$ \\ 
                                            & $\sigma/\mathrm{eV}$                                          & $88.1\err{21.0}{24.0}$ & $76.3\err{34.0}{29.0}$  \\ 
                                            & $\mathrm{Norm}/10^{-5}$                                       & $3.91\err{0.56}{0.65}$ & $3.79\err{0.52}{0.44}$  \\  \hline
    \end{tabular}
    \label{tab:model12}
\end{table*}

\begin{table}
\renewcommand{\arraystretch}{1.25}
    \centering
    \caption{Best fit values of parameters left free across all six phases. In model 1 (M1) these parameters are \textsc{zxipcf} covering fraction and \textsc{nthcomp} normalisations. In model 2 (M2) these are \textsc{nthcomp$_{\mathrm{hard}}$} power law index, and normalisation, and \textsc{nthcomp$_{\mathrm{soft}}$} normalisation.}
    \begin{tabular}{lccc}
    \hline
        & \textbf{\textsc{zxipcf}} & \textbf{\textsc{nthcomp$_{\mathrm{soft}}$}} & \textbf{\textsc{nthcomp$_{\mathrm{hard}}$}} \\
        \textbf{M1} & $f_{cov}$ & $\mathrm{Norm}/10^{-3}$ & $\mathrm{Norm}/10^{-3}$ \\ \hline\hline
        \textbf{L1} & $0.45\err{0.04}{0.03}$ & $2.17\err{0.22}{0.41}$ & $11.4\err{0.6}{0.8}$ \\
        \textbf{H1} & $0.29\err{0.06}{0.04}$ & $2.63\err{0.25}{0.48}$ & $13.3\err{0.8}{1.0}$ \\ 
        \textbf{L2} & $0.41\err{0.05}{0.03}$ & $1.95\err{0.21}{0.36}$ & $10.9\err{0.6}{0.8}$ \\ 
        \textbf{H2} & $0.32\err{0.06}{0.04}$ & $2.26\err{0.22}{0.39}$ & $11.9\err{0.7}{0.9}$ \\ 
        \textbf{L3} & $0.49\err{0.04}{0.03}$ & $1.72\err{0.19}{0.31}$ & $11.4\err{0.6}{0.8}$ \\
        \textbf{H3} & $0.29\err{0.06}{0.04}$ & $2.20\err{0.22}{0.40}$ & $11.5\err{0.7}{0.8}$ \\ \hline\\ \hline
        & \textbf{\textsc{nthcomp$_{\mathrm{hard}}$}} & \textbf{\textsc{nthcomp$_{\mathrm{soft}}$}} & \textbf{\textsc{nthcomp$_{\mathrm{hard}}$}} \\ 
        \textbf{M2} & $\Gamma$  & $\mathrm{Norm}/10^{-3}$ & $\mathrm{Norm}/10^{-3}$ \\ \hline\hline
        \textbf{L1} & $1.80\err{0.03}{0.04}$ & $2.09\err{0.38}{0.52}$ & $10.3\err{0.7}{0.9}$ \\ 
        \textbf{H1} & $1.86\err{0.03}{0.04}$ & $2.58\err{0.49}{0.66}$ & $13.8\err{1.0}{1.3}$ \\ 
        \textbf{L2} & $1.83\pm{0.04}$ & $1.88\err{0.34}{0.50}$ & $10.5\err{0.8}{0.9}$ \\
        \textbf{H2} & $1.90\pm{0.04}$ & $2.23\err{0.42}{0.60}$ & $13.3\err{1.0}{1.2}$ \\ 
        \textbf{L3} & $1.76\err{0.03}{0.04}$ & $1.66\err{0.33}{0.42}$ & $9.47\err{0.6}{0.9}$ \\ 
        \textbf{H3} & $1.92\err{0.03}{0.04}$ & $2.17\err{0.43}{0.57}$ & $13.5\err{0.9}{1.3}$ \\ \hline
    \end{tabular}
    \label{tab:freepar12}
\end{table}

\section{Discussion}
\label{sec:discussion}
\subsection{Detection of Periodic Behaviour in NGC~6814}
Three periods of cyclic behaviour are evident in the X-ray light curve of NGC~6814 prior to the eclipse. A superlet transform is used to quantify the behaviour and establish its significance.  The period is approximately $20$\,ks.  From a purely statistical sense, this period is only marginally significant (\textgreater90\% confidence) in the broad band ($0.3-12.0$\,keV) and soft ($0.3-1.0$\,keV) light curves, but there are spectral changes associated with the phase of the variability indicating the behaviour is not random, but associated with physical changes in the AGN  (Section~\ref{sec:spectral}). In comparison to \rej, our findings exhibit some similarities with \cite{ghosh2023} which found that the CWT of a single observation had 90\% significance across the entire observation with peaks of 95\% significance. The difference in the case of NGC~6814 is that the SLT power and significance is reduced during the eclipse (Fig.~\ref{fig:slt_stack}, top).

The superlet transform reveals the period of interest in all light curves and suggests the behaviour is still active during the eclipse (see Appendix \ref{sec:slt_appendix}).  Moreover, the superlet transform of the 2021 \textit{XMM-Newton} observation highlights the same frequency of interest, but with less significance (see Appendix \ref{sec:obs_appendix}).  This could be indicative of a long-lived periodic behaviour that is normally buried in the dominant red-noise variations associated with the AGN. This may be similar to the QPO in \rej, which has been found to be persistent across multiple observations over the course of more than a decade \citep[e.g.][]{taylor2025}.

\subsection{Modulation of the Signal and Possible Delays During the Eclipse}
As mentioned, the superlet transforms suggest the period persists during the eclipse. Closer inspection showed that the cyclic behaviour does exists during the eclipse, but that the period may be modulated or stretched (Section~\ref{sec:mod}) and the effect seems to be  independent of the energy band (Fig.~\ref{fig:normlc}).

One hypothesis considered was that the signal would be modified as the radiation transmitted through the obscuring medium of the eclipse by time-dependent photoionisation effects, due to the obscuring clouds responding to changes in the radiation. To test this, \textsc{tpho} \citep{rogantini2022} simulations of an intrinsically sinusoidal signal traversing an absorbing cloud were carried out.  The cloud properties covered the range in column density and covering fraction reported by \cite{pottie2023} for the NGC~6814 eclipse.  However, no modification to the input signal was observed.  The time-dependent photoionisation effects, at least at the column and gas densities considered here, do not have a significant impact on the input light curve.

Examination of the DCF prior to the eclipse reveals a significant correlation at zero delay between the low-energy light curve and the highest energy ($6.0-12.0$\,keV) light curve (Fig.~\ref{fig:dcf_stack}, top).  However, during the eclipse, the same DCF peaks at zero delay but exhibits a strong (\textgreater$2\,\sigma$) asymmetry to negative lags (Fig.~\ref{fig:dcf_stack}, bottom) indicating the unobscured hard band leads the obscured soft variability.  This could indicate that the travel path of some of the soft photons is longer than the unobscured hard photons.  This observation is consistent with the picture from \cite{pottie2023} and \cite{kang2023} that the eclipsing medium is not a single cloud, but rather it is clumpy and inhomogeneous.  In an X-ray colour analysis, \cite{pottie2023} demonstrated that the eclipsing medium was likely made up of multiple clouds and perhaps embedded in a highly ionised halo, and the distribution of these clouds was not isotropic. 

One can envisage a scenario where hard X-rays will cross such a medium and arrive at the observer unimpeded.  On the other hand, some low-energy photons may ``bounce’’ (i.e reflect or scatter) off such clouds, altering their path (i.e. distance travelled), before they arrive at the observer delayed to the hard X-rays.  Since the covering fraction is not 1, this will not affect all X-rays, and it would create an asymmetry in the DCF rather than a delayed peak.

While this explanation adequately describes the delay seen in the DCF during the eclipse, it does not explain how the period is modulated by a similar amount in all energy bands. If the eclipse does modulate the period, and the hard band was unaltered by the eclipse, we might expect to see hard band frequency remain constant throughout the observation. However, fitting the hard band with such a frequency ($\sim53$\,\muhz) results in a statistically poorer fit. Instead, it may be possible that some component of the eclipsing medium is responsible for modulating the period at all energies studied, while the majority of the clouds are responsible for scattering/absorbing only the soft photons and causing the observed delay between bands.

\subsection{Origin of Periodic Behaviour in NGC~6814}
The origin of the periodic behaviour in NGC~6814 is unclear. The measured period is approximately $6\times$ longer than the period seen in the prototypical AGN-QPO, \rej.  Since the central SMBH mass of NGC~6814 is $\sim10^7\,M_\odot$ \citep{bian2010} and that of \rej\ is $10^6\,M_\odot$ \citep{bentz2015}, it could point to a similar QPO origin with comparable time scales that simply scale with black hole mass.

For \rej, it has been suggested that the QPO originates within a few $R_g$ of the SMBH \citep{zoghbi2011, middleton2011}, and is attributed to fluctuations in the hot corona. \cite{czerny2012} suggests that QPOs may form through shock oscillations in low angular momentum hot accretion flows. In this case, the variability of the shock location results in the observed variability in X-ray flux and QPO period. \cite{taylor2025} suggest the QPO originates in the hot corona and the signal in the soft band is attributed to contamination from the hard power.  They also report some delay in the soft band as hot corona emission is reprocessed in the disc.   Here, a couple of attractive scenarios are that the inner truncated disc is replaced with a hot misaligned flow that is precessing \citep{bollimpalli2024, bollimpalli2023}, or that the corona is the base of a failed jet that might be misaligned with the disc \citep{taylor2025}.

In the case of NGC~6814, associating the period to a dynamical time scale places the source at a distance of $\sim17\,r_g$. \cite{gallo2021} found the X-ray emitting region to be about $25\,r_g$ across.  In addition, modeling of the optical-to-X-ray spectral energy distribution by \cite{pothier2025} \citep[see also][]{gonzalez2024} measured the inner accretion disc to be \textgreater$60\,r_g$. This would place the origin of the oscillation in the X-ray emitting region and not in the disc or warm corona, in agreement with the previous works on \rej.  This scenario appears broadly consistent with our Model 2 in Section~\ref {sec:spectral} that shows the bulk of the variability to be attributed to the hot corona changes in brightness and photon index.  In contrast, the lag-frequency and lag-energy spectra in NGC~6814 are effectively featureless, but the duration of this observation is extremely modest in comparison to \cite{taylor2025} that made use of more than a megasecond of data for \rej.

Alternatively, Model 1 in Section~\ref {sec:spectral} raises the possibility that the observed changes in brightness are from periodic changes in the level of obscuration (i.e. covering fraction) of a roughly constant intrinsic source.  Again, if this time scale is dynamical, the source of obscuration would be located at $\sim17\,r_g$, which can still be consistent with the truncated disc and large corona scenario \citep[e.g.][]{pothier2025, gonzalez2024, gallo2021}.  Conceivably, if the inner region is replaced with a hot, misaligned flow, a precessing annulus as per numerical simulations of warped and torn discs \citep[e.g.][]{Raj2021, Fragile2026}, could produce the obscuration of the primary source.

Given the association to the eclipsing event, it is difficult to rule out a line-of-sight absorption origin for this variability. In particular, the third trough (see L3 in  Table~\ref{tab:freepar12}) shows significant hardening, consistent with the obscuring material of the eclipse.  Moreover, the $\sim20$~ks cycle aligns with the ingress and egress of the eclipse (Figure~\ref{fig:phasefold}).  The changes in covering fraction in Model 1 could be a result of obscuring material moving through the line-of-sight, possibly associated with the obscuring clouds of the eclipsing region. The absorption interpretation for the difference spectrum also favours this scenario.  \cite{pottie2023} demonstrated that the eclipsing medium was made up of multiple clouds embedded in a highly ionised halo.  If clouds leading the primary eclipse are spaced out at roughly 20\,ks intervals, the period and eclipse could be induced by the material in the same region.

However, the modulation of the periodic signal during the eclipse would be better described if the period was intrinsic (e.g. to the corona) rather than an obscuration effect.

\section{Conclusions}
\label{sec:conclusions}
In this paper we have performed a combined timing and spectral analysis of NGC~6814 to uncover the characteristics of the periodic behaviour seen in the 2016 \textit{XMM-Newton} light curve. We developed a Python package for performing wavelet/superlet analysis of AGN light curves \citep{hodd2026}, including light curve simulation and significance estimation. The results of our timing and spectral analyses suggest two possible origins for the periodic behaviour.

\begin{enumerate}
  \item The superlet transform of the light curve confirms an oscillation at the 90\% significance level with a period of $22\err{3}{2}$\,ks in the first 60\,ks of the observation apparent at all energies, though with deminished significance in the hard band because of reduced signal-to-noise.  Fitting a sinusoid to the first 60\,ks gives a similar period and shows that the ingress and egress of the eclipse is in-phase with the oscillation. This periodic behaviour is also evident in the 2021 \textit{XMM-Newton} observation of NGC~6814 (Fig.~\ref{fig:slt_21}). Fourier frequency-lag analysis shows no significant lag features between energy bands.
  \item The discrete correlation functions between soft and hard bands support the detection of this periodic behaviour prior to the eclipse. Further, they show that during the eclipse the hard band leads the soft, which is consistent with a clumpy eclipsing medium.
  \item In our modelling, time-dependent photoionisation effects in the obscuring material are insufficient to modulate the frequency to the degree seen in light curves that have had the effects of the eclipse removed.
  \item Spectral analysis shows that the periodic behaviour results in changes to the spectral shape and normalisation. Two models can describe the data. First, that the periodic behaviour is caused by changes in the covering fraction of the obscuring material. This is supported by the difference spectrum and may be associated with the clumpy eclipsing clouds. Alternatively, the periodic behaviour is caused by intrinsic changes in the hot corona or in a truncated disc replaced with a misaligned flow.  This scenario appears consistent with other analyses of NGC~6814 that report truncated discs and compact coron\ae.
\end{enumerate}

While it is still difficult to state the exact cause of the periodic behaviour in the X-ray light curves of NGC~6814, this work identifies the two most probable causes as either inhomogeneous obscuring material possibly associated with the eclipse, or by some changes in the inner accretion flow. Deeper observations are needed to confirm and investigate the QPO behaviour in NGC~6814.


\begin{acknowledgments}
We thank the referee and scientific editors for their helpful comments that improved the manuscript. LCG acknowledges support from the Natural Sciences and Engineering Research
Council of Canada (NSERC) and the Canadian Space Agency (CSA).
\end{acknowledgments}




%
\facilities{XMM-Newton (EPIC-pn), Suzaku (XIS)}

\software{\texttt{XSuperlet} \citep{hodd2026},
          \texttt{pyLag} (\url{github.com/wilkinsdr/pyLag}),
          \texttt{wwz} \citep{kiehlmann2023},
          \texttt{pyDCF} \citep{robertson2015},
          \textsc{xspec} \citep{arnaud1996},
          \textsc{spex} \citep{kaastra1996},
          \texttt{numpy} \citep{numpy2020},
          \texttt{scipy} \citep{scipy2020},
          \texttt{scikit-learn} \citep{scikit-learn},
          \texttt{matplotlib} \citep{hunter2007}
          }



\appendix

\section{Additional information on the superlet transform}
\label{sec:slt_appendix}
In the Fourier transform, the input signal is decomposed into a sum of sinusoidal components. These sinusoids are time invariant and span the entire time domain, thus the Fourier transform is an average for the entire time series, and no time domain information is kept. This results in an implicit assumption of a stationary process where the properties of variability are assumed constant which may not be the case for X-ray light curves of AGN \cite[e.g.][]{alston2019}.

Since the X-ray emitting region of an AGN is known to be highly variable we may want to use a technique that does not assume a stationary process and offers some time domain information. In such a case, the wavelet transform may be used in place of the Fourier transform in our analysis. Instead of the time invariant sinusoids of the Fourier transform, wavelet analysis relies on time localised wave packets (wavelets). The field of wavelet analysis covers a huge number of applications across the sciences and engineering. Wavelet analysis has found many uses in medicine including analysis of electrocardiograms (ECGs) and electroencephalograms (EEGs). It has been used in data analysis in meteorology and climate science \citep{torrence1998}, and in fluid dynamics and non-destructive testing in engineering. It is not unfamiliar in X-ray astronomy either, having been used to analyse time variability of QPOs \citep{czerny2012, ghosh2023}, X-ray binaries \citep{lachowicz2005}, and in spectral timing \citep{ghosh2023} and X-ray reverberation of AGN \citep{wilkins2023}.

By combining multiple wavelet transforms, good time and frequency resolution can be achieved simultaneously using a transform introduced by \cite{moca2021} and dubbed the superlet transform. This type of transform has been found to work well on synthetic data and brain signals since it is able to resolve fast and transient signals with precision. It has quickly found uses in medicine \citep{singhal2024}, in signal processing \citep{mignone2023}, and even in the acoustic testing of bridges \citep{hu2025}. Elsewhere in physics, it has been used to resolve picosecond timescale signals in quantum mechanics \citep{gentry2022}, and in molecular dynamics \citep{feng2025}.

The superlet transform may be calculated using either additive or multiplicative wavelets. This affects the maximum number of cycles, and the spacing of cycles in the intermediate wavelets. With additive wavelets: $c_i=c_1+i-1$, with multiplicative wavelets: $c_i=c_1\cdot i$. \textsc{XSuperlet} defaults to multiplicative wavelets which are used throughout this paper.

The following figures (Fig.~\ref{fig:slt_soft}~-~\ref{fig:slt_hard}) are superlet scalograms for each band investigated.

\begin{figure}
	\includegraphics[width=\columnwidth]{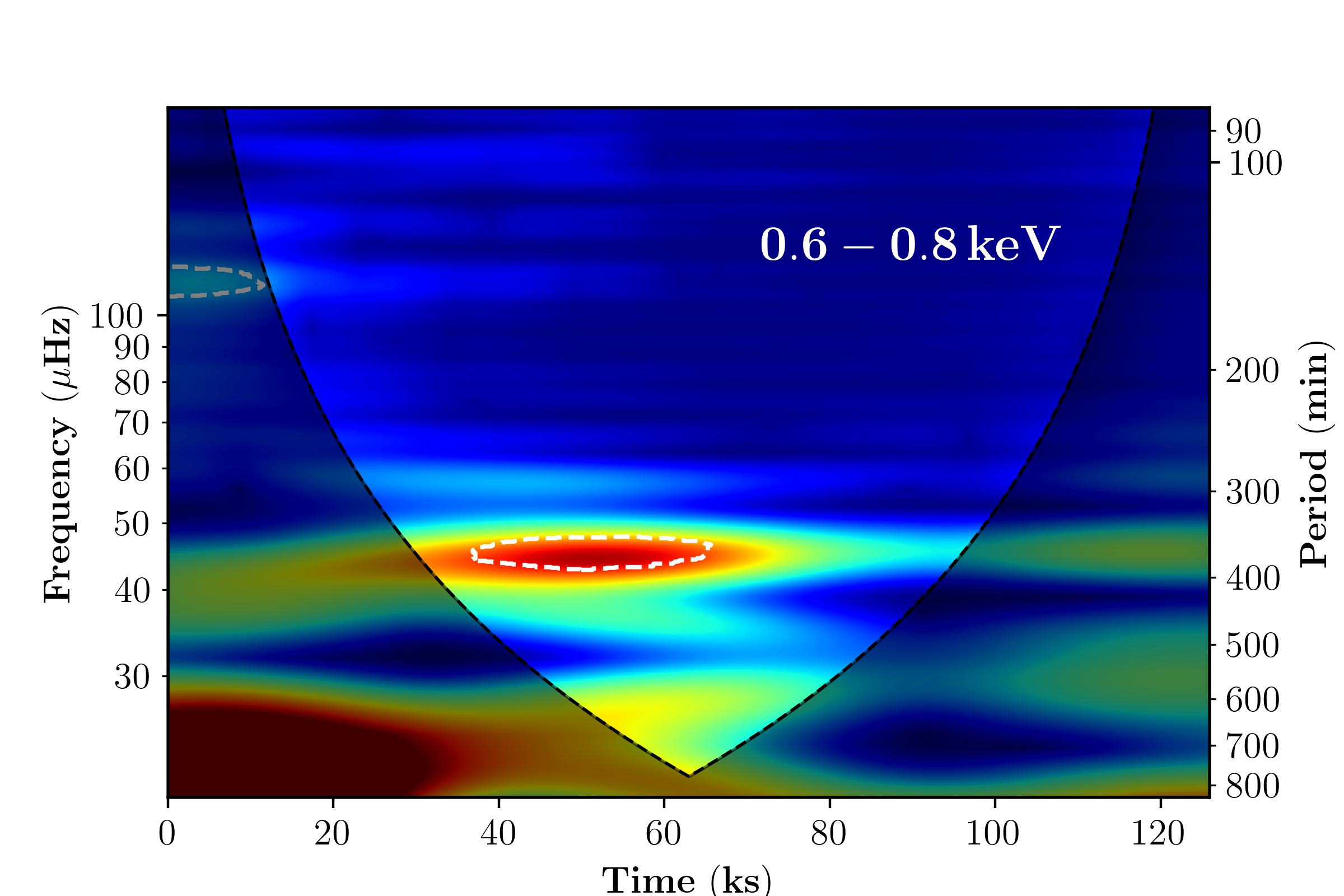}
    \caption{SLT for the very soft band $0.6-0.8$\,keV. Dashed lines show the 90\% confidence regions. The dark shaded regions denote areas outside COI.}
    \label{fig:slt_soft}
\end{figure}

\begin{figure}
	\includegraphics[width=\columnwidth]{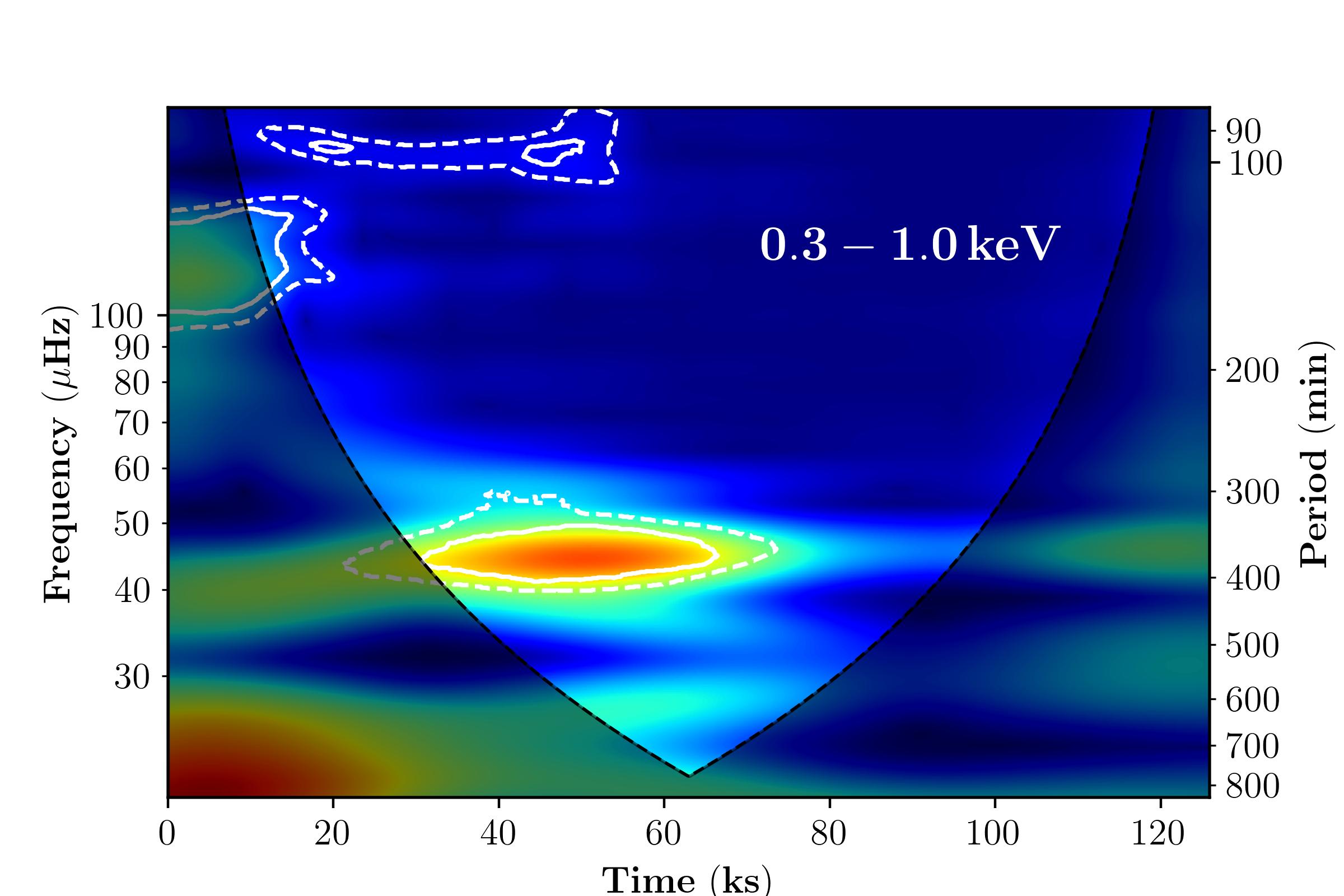}
    \caption{SLT for the soft band $0.3-1.0$\,keV. Dashed lines show the 90\% confidence regions and solid lines show the 95\% regions. The dark shaded regions denote areas outside COI.}
    \label{fig:slt_soft_p31}
\end{figure}

\begin{figure}
	\includegraphics[width=\columnwidth]{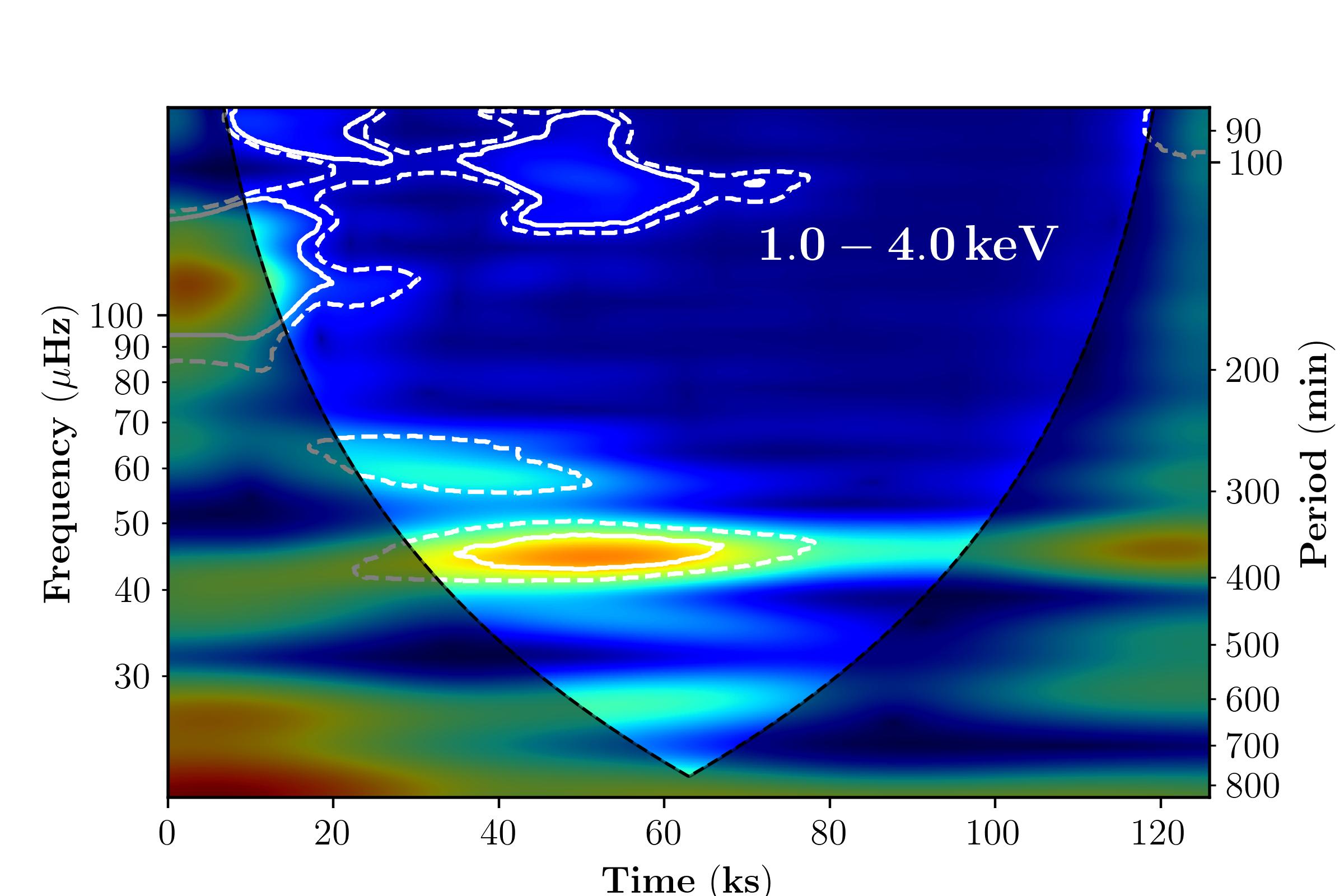}
    \caption{SLT for the mid band $1.0-4.0$\,keV. Dashed lines show the 90\% confidence regions and solid lines show the 95\% regions. The dark shaded regions denote areas outside COI.}
    \label{fig:slt_mid}
\end{figure}

\begin{figure}
	\includegraphics[width=\columnwidth]{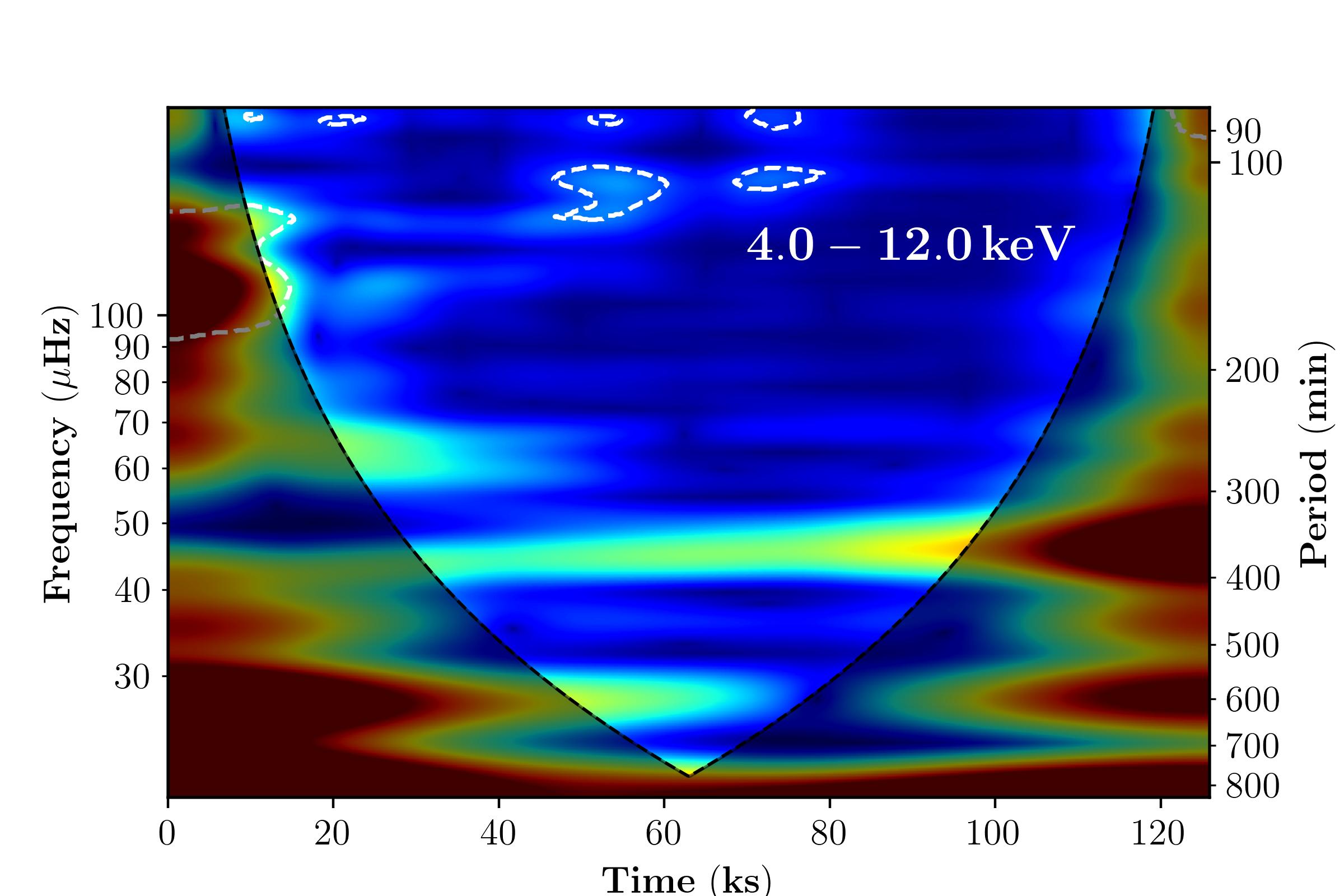}
    \caption{SLT for the hard band $4.0-12.0$\,keV. Dashed lines show the 90\% confidence regions. The dark shaded regions denote areas outside COI.}
    \label{fig:slt_hard}
\end{figure}

\section{Wavelet transforms of other observations}
\label{sec:obs_appendix}
The following figures (Fig.~\ref{fig:slt_21}~-~\ref{fig:wwa_11}) are wavelet scalograms for the 2021 \textit{XMM-Newton} and 2011 \textit{Suzaku} observations. Note that since the \textit{Suzaku} light curve contains gaps we use the weighted wavelet z-transform (WWZ, \cite{foster1996}) in the place of the SLT which requires continuous and evenly sampled data. WWZ transforms are calculated through \textsc{XSuperlet} using \textsc{wwz.py} \citep{kiehlmann2023}.

\begin{figure}
	\includegraphics[width=\columnwidth]{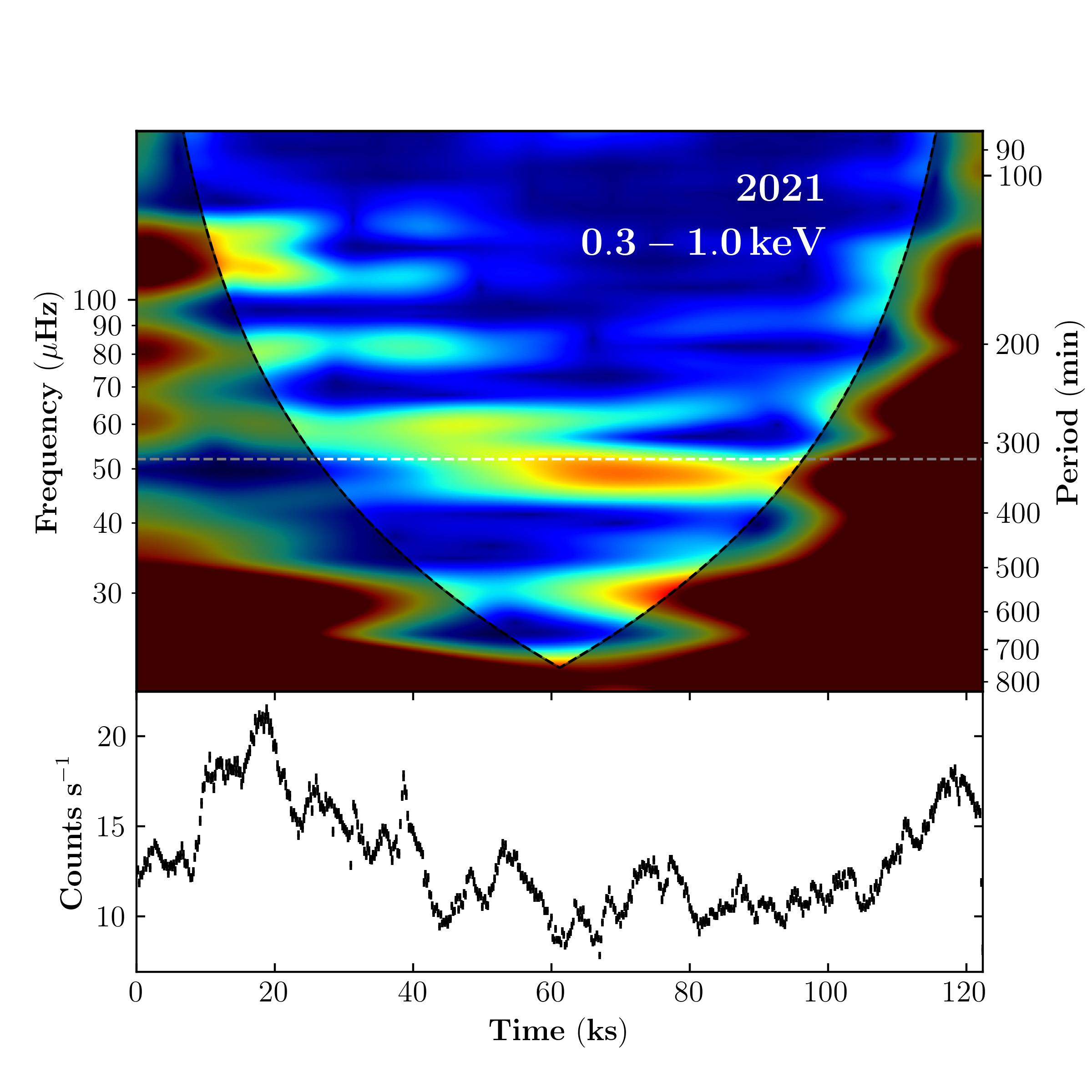}
    \caption{SLT for the broad band $0.3-10.0$\,keV 2021 \textit{XMM-Newton} observation. The white dashed line is at the $52$\muhz\ frequency found with the sine fit in Section~\ref{sec:timing}. The dark shaded regions denote areas outside COI.}
    \label{fig:slt_21}
\end{figure}

\begin{figure}
	\includegraphics[width=\columnwidth]{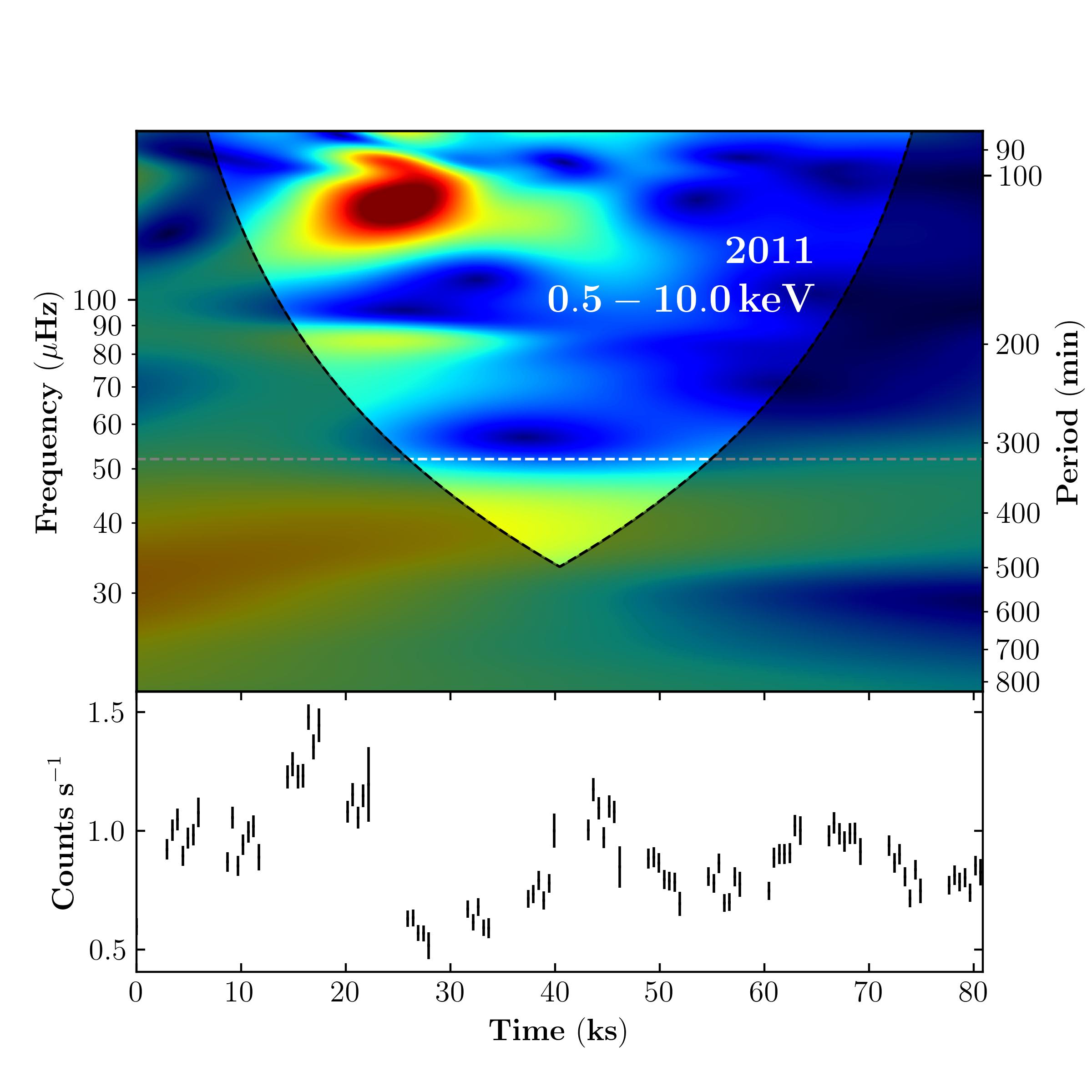}
    \caption{Weighted wavelet amplitude (WWA) for the broad band $0.5-10.0$\,keV 2011 \textit{Suzaku} observation. The white dashed line is at the $52$\muhz\ frequency found with the sine fit in Section~\ref{sec:timing}. The dark shaded regions denote areas outside COI.} 
    \label{fig:wwa_11}
\end{figure}

\section{Power spectra}
\begin{figure}
	\includegraphics[width=\columnwidth]{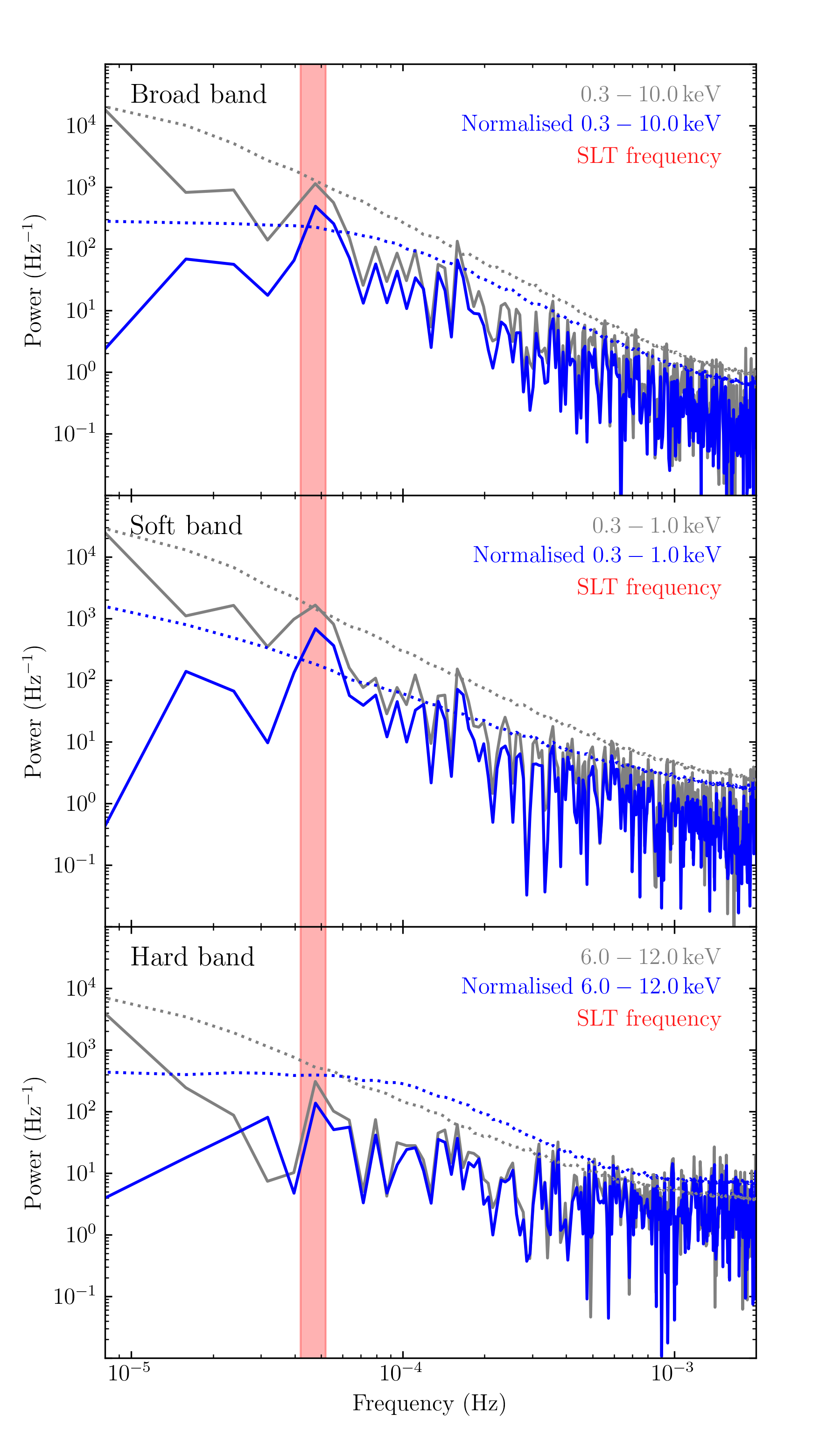}
    \caption{PSDs for the broad band $0.3-10.0$\,keV, soft band $0.3-1.0$\,keV, and very hard band $6.0-12.0$\,keV light curves. The region highlighted in red shows the frequency detected by SLT. In blue are the PSDs of the normalised light curves. The dotted lines denote the 90\textsuperscript{th} percentile based on the PSDs of 1000 simulated light curves for each band.}
    \label{fig:psd_broad}
\end{figure}

Here we present power spectra (periodograms) for several light curves used in this work (Fig.~\ref{fig:psd_broad}).


\bibliography{Bib}{}
\bibliographystyle{aasjournalv7}



\end{document}